\documentclass[aps,prb,twocolumn,floatfix,amsmath,showpacs]{revtex4}
\usepackage{graphicx}

\newcommand{\bd}{\begin{displaymath}}
\newcommand{\ed}{\end{displaymath}}
\newcommand{\beq}{\begin{equation}}
\newcommand{\eeq}{\end{equation}}

\begin{document}
\title{ Short-range interactions in a two-electron system: energy
levels and magnetic properties }

\date{ \today}

\author{L.G.G.V. Dias da Silva$^{1,2.*}$, M.A.M. de Aguiar$^{1}$}
\affiliation{$^1$ Instituto de F\'{\i}sica `Gleb Wataghin',
Universidade Estadual de Campinas (UNICAMP), \\
Caixa Postal 6165, 13083-970 Campinas, Brazil}
\affiliation{$^2$ Duke University - Physics Department, \\
P.O. Box 90305 - Durham, NC  27708-0305 USA}

\begin{abstract}

The problem of two electrons in a square billiard interacting via a
finite-range repulsive Yukawa potential and subjected to a constant
magnetic field is considered. We compute the energy spectrum for both
singlet and triplet states, and for all symmetry classes, as a
function of the strength and range of the interaction and of the
magnetic field. We show that the short-range nature of the potential
suppresses the formation of ``Wigner molecule'' states for the ground
state, even in the strong interaction limit. The magnetic
susceptibility $\chi(B)$ shows low-temperature paramagnetic peaks due
to exchange induced singlet-triplet oscillations. The position, number
and intensity of these peaks depend on the range and strength of the
interaction. The contribution of the interaction to the susceptibility
displays paramagnetic and diamagnetic phases as a function of $T$.
\end{abstract}


\maketitle



\section{Introduction}

The study of mesoscopic systems has proved to be a rich field to
investigate explicit manifestations of quantum properties in nanometer
and micron scales \cite{Mesosc_review}. In such systems, the electron
coherence length scales and mean-free paths are in general larger than
the typical sample sizes, so that the underlying classical electronic
motion plays an important role. The nature of the classical motion,
regular, mixed or chaotic, reflects itself on some of the quantum
properties of the system, particularly in the energy level
distribution. These features have long been studied for
non-interacting \cite{chaos_nonint} and weakly interacting systems
\cite{harold}.

In quantum dots, where few electrons are laterally confined
\cite{kastner}, the electron-electron interaction
is usually very efficiently shielded by the positively charged fixed
ions and other effects, so that the independent electron gas theory
can often be used to understand the basic features of the system
\cite{ullmo,report,levy,prado1,prado,agam,oppen}. However, the
residual interaction that survives the shielding can sometimes play
important roles. In recent years, much attention has been given to
interaction-induced effects in mesoscopic systems
\cite{bedanov,peeters,harold,creffield1,creffield2,grigorenko,akbar}.
These effects are particularly important in the large-dot regime,
where the electron-electron Coulomb interaction overcomes the kinetic
energy, forcing the ground-state into a {\it Wigner molecule} type of
configuration \cite{creffield1}.

In mesoscopic systems the electronic interaction is usually not well
approximated by a bare long-range Coulomb force, due exactly to
screening effects \cite{ando}. The strength and range of the residual
interaction, or the efficiency of the screening, depend on many
parameters, like the electron density and size of the dot. It is
therefore important to understand the effects of the interaction as a
function of its effective intensity and range. In this work we give a
contribution in this direction presenting exact results for the
problem of two electrons in a square quantum dot interacting via a
repulsive finite-range Yukawa type of interaction, $V(r)=V_0
e^{-\alpha r}/r$, and subjected to a uniform and constant magnetic
field of strength $B$ applied perpendicular to the dot. This model
system was inspired by the experimental work of Levy et al \cite{levy}
where the orbital magnetic susceptibility was measured for an ensemble
of square dots containing of the order of a thousand electrons
each. Although the independent particle semiclassical theory explained
most of the experimental findings
\cite{ullmo,report,levy,prado1,prado,agam,oppen}, the behavior of the
susceptibility with the temperature does not come out correctly in
this approach and still puzzles the theorists. The idea that the slow
decay of the susceptibility observed experimentally (as opposed to the
exponential decay expected from the semiclassical theory), could be
due to electron-electron interaction was first investigated in
\cite{harold} for a weak contact (Dirac delta) type of interaction
using perturbation methods. In this article we study the effects of
electronic interaction a much simpler system, with only two electrons,
but we present exact (numerical) results.

The choice of an Yukawa type of potential allows us to interpolate
between the pure Coulomb ($\alpha=0$) and short range
interactions. Besides, the calculation of the Hamiltonian matrix
elements can be reduced to one-dimensional integrals, which can be
calculated numerically. This allows us to compute the energy spectrum
for the four rotational symmetry classes as a function of the
interaction strength $V_0$ and range $1/\alpha$ for both singlet and
triplet states. We also consider these results as a function of a
constant magnetic field of strength $B$ applied perpendicular to the
square. We compute the magnetic susceptibility at finite values of the
magnetic field and temperature via the partition function.

Our main results can be summarized as follows: (1) $V_0$ introduces
avoided crossings between the energy levels within each symmetry
class, one of the signatures of quantum chaos; (2) $\alpha$ has a very
important role in determining the probability profile of the ground
state, suppressing in some cases the Wigner molecule type of behavior
even for strong interactions; (3) The effect of the interaction on the
magnetic susceptibility $\chi(B)$ depends on $\alpha$. In particular,
for large magnetic fields, singlet-triplet oscillations of the ground
state level lead to paramagnetic fluctuations on the two-electron
susceptibility $\chi(B)$, in contrast to the non-interacting
diamagnetic susceptibility. The position and intensity of these peaks
change with the range of the interaction.; (4) The contribution to the
susceptibility induced by interaction at zero magnetic field shows
paramagnetic and diamagnetic phases as a function of the
temperature. This type of behavior has also been found for weak Dirac
delta interactions \cite{harold}.

This paper is organized as follows: in section II we describe the
system in detail. We discuss its symmetry properties and compute the
matrix elements of the Hamiltonian. In section III we present
numerical results for the energy spectrum and for ground-state
electronic density as a function of the strength and range of the
interaction. In section IV we consider the magnetic properties of the
system and in section V we discuss our results.

\section{Hamiltonian and Matrix Elements}

We consider a system where two electrons are confined in a
square-shaped two dimensional billiard of size $L$ interacting via an
Yukawa type of potential and subjected to a uniform and constant
magnetic field of strength $B$ applied perpendicular to the dot. The
Hamiltonian is given by
\begin{eqnarray}
 H = && \frac{1}{2m^{*}} \sum_{i=1,2} (p_{xi}+eBy_i/2)^2  +
    (p_{yi}-eBx_i/2)^2 )  \nonumber\\
&&+  V_{walls} + V_0 \frac{e^{-\alpha |\vec{r}_1-\vec{r}_2|}}{|\vec{r}_1-\vec{r}_2|}
\label{hamilto}
\end{eqnarray}
where $m^{*}$ is the quasiparticle electron mass and $V_{walls}$ is
the square well potential.
The eigenfunctions of a single particle in this square dot with zero
magnetic field are given by a normalized product of sine functions:
\begin{equation}
\varphi_{m n} (x,y) = \left( \frac{2}{L} \right) \sin{\frac{ m \pi }{L} x} \sin{\frac{ n \pi }{L} y}
\label{fi2d}
\end{equation}
and the eigenenergies are simply
\begin{equation}
E_{mn} = \frac{\hbar^2}{2 m^{*} L^2} \pi^2 (m^2 + n^2)
\label{E2d}
\end{equation}
The square billiard is a highly symmetric system. It is invariant
under the action of 8 symmetry operations (4 rotations plus 4
reflections) which form the ${\mathcal C}_{4v}$ symmetry group.  When
the time-reversal symmetry is broken (e.g. by the application of a
magnetic field), the Hamiltonian is no longer invariant under
reflections. The group then reduces to ${\mathcal C}_{4}$, formed by
the four rotations generated by $\hat{C}_4$ (rotation by $\frac{\pi}{2}
$ ). The symmetric eigenfunctions can be written as a linear
combination of a particular eigenfunction and its symmetry-related
counterparts:

\begin{equation}
\psi(x,y) = \varphi(x,y) + \hat{C_4} \varphi(x,y) + \hat{C^{2}_4} \varphi(x,y) + \hat{C^{3}_4} \varphi(x,y) \;.
\label{comblin}
\end{equation}

Rotating $\psi$ leads to $\hat{C_4} \psi = e^{i \theta} \psi$ with
$(e^{i \theta})^4 = 1$. This, in turn, leads to four solutions for
$e^{i \theta}$, namely $+1,-1,+i$ and $-i$. We can thus separate the
general eigenfunctions (\ref{fi2d}) in four ``classes'' (or
representations) using the group's character table \cite{hammermesh},
as follows:

\begin{equation}
\begin{array}{rcl}
\psi_{mn}^{(+1)}(x,y) & = &  \left\{ \begin{array}{lll} \varphi_{mn}(x,y) & &
\mbox{\small if $n = m$ (both odd)} \\
\multicolumn{3}{l}{ \frac{1}{\sqrt{2}} \left(\varphi_{mn}(x,y) (\pm)  \varphi_{nm}(x,y) \right) } \\
&& \mbox{\small if $n \neq m$ and both odd (even)}
\end{array} \right. \\[0.1cm]
\psi_{mn}^{(-1)}(x,y) & = &  \left\{ \begin{array}{lll} \varphi_{mn}(x,y) & &
\mbox{\small if $n = m$ (both even)} \\
\multicolumn{3}{l}{ \frac{1}{\sqrt{2}} \left(\varphi_{mn}(x,y) (\mp)  \varphi_{nm}(x,y) \right) } \\
&& \mbox{\small if $n \neq m$ and both odd (even)}
\end{array} \right. \\[0.1cm]
\psi_{mn}^{(+i)}(x,y) & = &  \left\{ \begin{array}{lll} \frac{1}{\sqrt{2}} \left(\varphi_{mn}(x,y)
\right. & \pm & i\varphi_{nm}(x,y) ) \\
& \multicolumn{2}{l}{\mbox{\small if  $m$ even (odd) $n$ odd (even)} } \end{array} \right. \\[0.1cm]
\psi_{mn}^{(-i)}(x,y) & = &  \left\{ \begin{array}{lll} \frac{1}{\sqrt{2}} \left(\varphi_{mn}(x,y)
\right. & \mp & i\varphi_{nm}(x,y) ) \\
& \multicolumn{2}{l}{\mbox{\small if  $m$ even (odd) $n$ odd (even)} } \end{array} \right. \\[0.1cm]
\end{array}
\label{psis}
\end{equation}

This equation can be written in a more compact form as
\begin{equation}
\psi^{(C)}_l = F^{(C)}_l \left( \varphi_l + S^{(C)}_l
\varphi_{\bar{l}} \right)
\label{psiscomp}
\end{equation}
where $l \equiv (m,n)$ and $\bar{l} \equiv (n,m)$. $F^{(C)}_l$ is
either $1$ or $\frac{1}{\sqrt{2}}$ and $S^{(C)}_l$ is $0$,
$\pm 1$ or $\pm i$, depending on the symmetry class $(C)$ and on $l$
(whether $(m,n)$ is odd/even and whether $m=n$ or $m \neq n$).

Finally the two-particle orbital eigenfunctions are symmetrized $(S)$
or anti-symmetrized $(A)$ combinations of one-particle orbital
eigenfunctions:
\begin{eqnarray}
\lefteqn{  \psi^{(S,A),(C_1 \otimes C_2)}_{l_1 l_2} (\vec{r_1},\vec{r_2})= }
\nonumber \\
& & \frac{1}{\sqrt{2}} \left( \psi_{l_1}^{(C_1)}(\vec{r_1}) \psi_{l_2}^{(C_2)}(\vec{r_2}) \pm
\psi_{l_1}^{(C_1)}(\vec{r_2}) \psi_{l_2}^{(C_2)}(\vec{r_1}) \right) \nonumber \\
\label{psi2p}
\end{eqnarray}
The orbital eigenfunction is symmetrized if the electrons are in
the singlet spin state and anti-symmetrized if they are in the triplet
spin state.

The symmetry group of the two-particle system is ${\mathcal C}_4 \otimes {\mathcal C}_4$ and the
eigenfunctions still separate in four symmetry classes. The two-particle ($2p$) class is defined by
the total phase gained under the action of an element of the ${\mathcal C}_4 \otimes {\mathcal
C}_4$ group ($E \otimes \hat{C}_4$, $\hat{C}_4 \otimes \hat{C}_4$ and so on). This phase is simply
the product of the one-particle ($1p$) phases in the representation shown in (\ref{psis}). The $2p$
class is thus obtained in a simple manner by ``multiplying'' the $1p$ classes. For instance, two
$1p$ states of class $(-1)$ form a $2p$ state of class $(+1)$ [pictorically, $(+1)$ "=" $(-1)
\otimes (-1)$]. The same happens with a $1p$ state of class $(+i)$ combined with other from class
$(-i)$. On the other hand, two $1p$ $(+i)$ states form a $(-1)$ $2p$ state and so on.
%
%

\subsection{ The screened Coulomb interaction}

For the electron-electron interaction we have used an ``Yukawa type''
short-range screened Coulomb potential
\begin{equation}
V(\vec{r_1},\vec{r_2}) = \frac{e^2}{4 \pi \epsilon_0 \epsilon_r}
\frac{e^{-\alpha |\vec{r_1}-\vec{r_2}|}}{|\vec{r_1}-\vec{r_2}|}
\label{pot}
\end{equation}
where $1/\alpha$ is the interaction range and $\epsilon_r$ is the
dielectric constant of the two-dimensional electron gas (in case of a
GaAs 2DEG, $\epsilon_r=10.9$). The reason for this particular choice
of screening is two-fold. First it interpolates between the pure
Coulomb case and localized interactions. Also it gives an effective
``interaction length'' ($\alpha^{-1}$) which is easy to
control. Second, the $1/r$ dependence is maintained with the screening
appearing as an exponential (as opposed to a power of $1/r$). This
facilitates enormously the calculation of the matrix elements, as we
show in the appendix. The range $\alpha$ will be considered here as a
free parameter.

The Hamiltonian for the two electrons without the magnetic field is
given by
\begin{equation}
\hat{H} = \sum_{i=1}^{2} \left( \frac{1}{2m_{i}^{*}}  \vec{p_i}^2 \right)
+  V(\vec{r_1},\vec{r_2}) \;.
\label{hamiltBV}
\end{equation}

Since the kinetic energy (\ref{E2d}) scales with $1/L^2$ and the
interaction scales with $1/L$, the electron-electron interaction term
dominates over the kinetic term for large $L$. Thus, we define an
``effective potential strength'' $V_0$ that grows linearly with the
dot size $L$:

\begin{equation}
V_0 \equiv L \frac{e^2}{4 \pi \epsilon_0 \epsilon_r}
\left( \frac{2 m^{*} }{ \hbar^2 \pi^2} \right)
\label{v0}
\end{equation}
so that we can write the matrix elements of $\hat{H}$ in the
non-interacting basis in units of $(\pi^2 \hbar^2)/(2 m^{*} L^2)$:
{\small
\begin{equation}
<\psi^{(C)}_{l_1 l_2} | \hat{H} | \psi^{(C)}_{l_1 l_2} >  =
\frac{\pi^2 \hbar^2}{2 m^{*} L^2} \left\{ \bar{E}_{l_1 l_2}   +
V_0 <\psi^{(C)}_{l_1 l_2} | \hat{V}(r/L) | \psi^{(C)}_{l_1 l_2} >
\right\}
\end{equation}
} where $\bar{E}_{l_1 l_2}=(m_{1}^2 + n_{1}^2 + m_{2}^2 + n_{2}^2)$ is the kinetic energy in units
of $(\pi^2 \hbar^2)/(2 m^{*} L^2)$. This defines another free parameter, $V_0$ (or, equivalently,
$L$) that controls the relative interaction strength.

The next step is to calculate the matrix elements of the potential in
the two-particle eigenfunction basis defined by Eq.(\ref{psi2p}). The
repulsive potential does not break the ${\mathcal C}_4 \otimes
{\mathcal C}_4$ rotational symmetry of the Hamiltonian, since it
depends only on the distance between the electrons. Therefore, the
interaction matrix is block-diagonal in this representation, i.e. the
matrix elements
\begin{equation}
V_{l_1 l_2 l^{'}_1 l^{'}_2} = \left<
\psi^{S,A,(\mbox{\scriptsize{Class}})}_{l_1 l_2} | \hat{V} |
\psi^{S,A,(\mbox{\scriptsize{Class}})}_{l^{'}_1 l^{'}_2} \right>
\label{pot_mat_element}
\end{equation}
are non-zero only inside the same symmetry block. For totally
symmetric (antisymmetric) eigenfunctions $V_{l_1 l_2 l^{'}_1 l^{'}_2}$
breaks into a sum (difference) of the a direct and an exchange term.
The general expression for the $\hat{V}$ matriz elements is given in
terms of the general one-particle states Eq.(\ref{psiscomp}) as
\begin{widetext}
\begin{equation}
\begin{array}{l}
\begin{array}{rcl}
<\psi_1^{A} \psi_2^{B}| \hat{V} |\psi_{1^{'}}^{C} \psi_{2^{'}}^{D}> &
= & F^{(A)}_1 F^{(B)}_2 F^{(C)}_3 F^{(D)}_4 \times \\[0.5cm]
\end{array} \\
\begin{array}{l}
\left( (1+ S^{(A)}_1 S^{(B)}_2 S^{(C)}_{1^{'}} S^{(D)}_{2^{'}})
<l_1 l_2 | \hat{V} |l^{'}_{1} l^{'}_{2} > +
(S^{(A)*}_1 S^{(B)*}_2 S^{(C)*}_{1^{'}} +
S^{(D)}_{2^{'}})<l_1 l_2 | \hat{V} |l^{'}_{1} \bar{l}^{'}_{2} >
\right.  \\
 + (S^{(A)*}_1 S^{(B)*}_2 S^{(D)*}_{2^{'}} + S^{(C)}_{1^{'}})
<l_1 l_2 | \hat{V} |\bar{l}^{'}_{1} l^{'}_{2} > +
(S^{(A)*}_1 S^{(C)*}_{1^{'}} S^{(D)*}_{2^{'}} +
S^{(B)}_2) <l_1 \bar{l}_2 | \hat{V} |l^{'}_{1} l^{'}_{2} >\\
 + (S^{(B)*}_2 S^{(C)*}_{1^{'}} S^{(D)*}_{2^{'}} + S^{(A)}_1)
<\bar{l}_1 l_2 | \hat{V} |l^{'}_{1} l^{'}_{2} > +
(S^{(A)*}_1 S^{(B)*}_2  + S^{(C)}_{1^{'}} S^{(D)}_{2^{'}})<l_1 l_2 |
\hat{V} |\bar{l}^{'}_{1} \bar{l}^{'}_{2} >\\
 \left. + (S^{(A)*}_1 S^{(C)*}_{1^{'}}  + S^{(B)}_2 S^{(D)}_{2^{'}})
<l_1 \bar{l}_2 | \hat{V} |l^{'}_{1} \bar{l}^{'}_{2} > +
(S^{(B)*}_2 S^{(C)*}_{1^{'}}  + S^{(A)}_1 S^{(D)}_{2^{'}})
<\bar{l}_1 l_2 | \hat{V} |l^{'}_{1} \bar{l}^{'}_{2} > \right) \;.\\
\end{array} \\
\end{array}
\label{elemgeral}
\end{equation}
The terms $<l_1 l_2 | \hat{V} |l^{'}_1 l^{'}_2 >$ can be written
explicitly with help of Eq.(\ref{fi2d}) as
{\small \begin{equation}
\left< l_1 l_2 | \hat{V} |l^{'}_1 l^{'}_2 \right>  = \frac{16}{L^4} \int \sin{\frac{\pi m_1
x_1}{L}} \sin{\frac{\pi n_1 y_1}{L}} \sin{\frac{\pi m_2 x_2}{L}} \sin{\frac{\pi n_2 y_2}{L}} \;
V(|\vec{r_2}-\vec{r_1}|)
 \; \sin{\frac{\pi m_1^{'} x_1}{L}} \sin{\frac{\pi n_1^{'} y_1}{L}}
\sin{\frac{\pi m_2^{'} x_2}{L}} \sin{\frac{\pi n_2^{'} y_2}{L}} \; d^2 \vec{r_1} d^2 \vec{r_2}
\label{Integral}
\end{equation} }
\end{widetext}
Eq.(\ref{elemgeral}) can be further simplified using the property
$\left< l_1 l_2 \right| \hat{V} \left| l^{'}_1 l^{'}_2 \right> =
\left< \bar{l}_1 \bar{l}_2 \right| \hat{V} \left| \bar{l}^{'}_1
\bar{l}^{'}_2 \right>$. The integrals in Eq.(\ref{Integral}) can be
evaluated by switching to relative polar $(r,\theta)$ and
center-of-mass coordinates. Thanks to the exponential form of the
screening, three of the four integrals in (\ref{Integral}) can be done
analytically. The remaining integral, over the relative polar angle
$\theta$, is performed numerically. Most of the direct and exchange
elements involve less than 16 integrals. The number is actually
$\frac{16}{2^N}$ where $N$ is the number of one-particle states with
$m=n$ involved in either one of the two-particle functions. All these
facts reduce the number of numerical integrals to be evaluated. The
details of this calculation are on Appendix A.

\subsection{ The magnetic field}

For $B\neq 0$, there are additional terms in the kinetic matrix
element of (\ref{hamilto}) proportional to $B$ and $B^2$. These terms
lead to integrals combining sine and cosine functions and powers of
$x$ and $y$, which can all be done analytically.

The terms linear in $B$ (involving $\hat{p}_y$ and $\hat{p}_y$)
contribute imaginary parts to the matrix elements, breaking the
degeneracy of the $(+i)$ and $(-i)$ symmetry classes. This is a
consequence of the time-reversal symmetry breaking.


\section{Effects of interaction: strength and range}

In this section, we show the numerical results obtained with the exact
diagonalization of the two-particle interacting Hamiltonian without
the magnetic field. We discuss the effects of the two independent
parameters of our model: the strength $V_0$ and the range $\alpha$.
To change the intensity of the interaction relative to the kinetic
energy we need to change $L$ (see Eq.(\ref{v0})). However, changing
$L$ changes the energy levels even if $V_0=0$. Therefore, in order to
focus on the changes induced only by the potential, we shall measure
the energy in units of $\frac{\pi^2 \hbar^2}{2 m^{*} L^2}$ throughout
this section. In these units, the non-interacting eigen-energies are
independent of $L$; the ground state energy, in particular, is equal
to 4.

\subsection{Energy levels}
\label{enlevels}

We first consider the effects of the interaction strength $V_0$. To
increase $V_0$ relative to the kinetic energy we need to increase
$L$. That, however, decreases the effective range of the
interaction. To keep the ratio between range and size of the dot
fixed, and concentrate on the effects of the potential strength, we
shall keep the {\it relative range} $1/(\alpha L)$ fixed as $L$ (or
$V_0$) is changed. When $1/(\alpha L) >1$, the electrons ``feel'' the
presence of each other all over the dot. For $1/(\alpha L) < 1$, the
interaction is more localized and the interaction range reduced.

\begin{figure*}
\includegraphics[height=5cm,width=5cm]{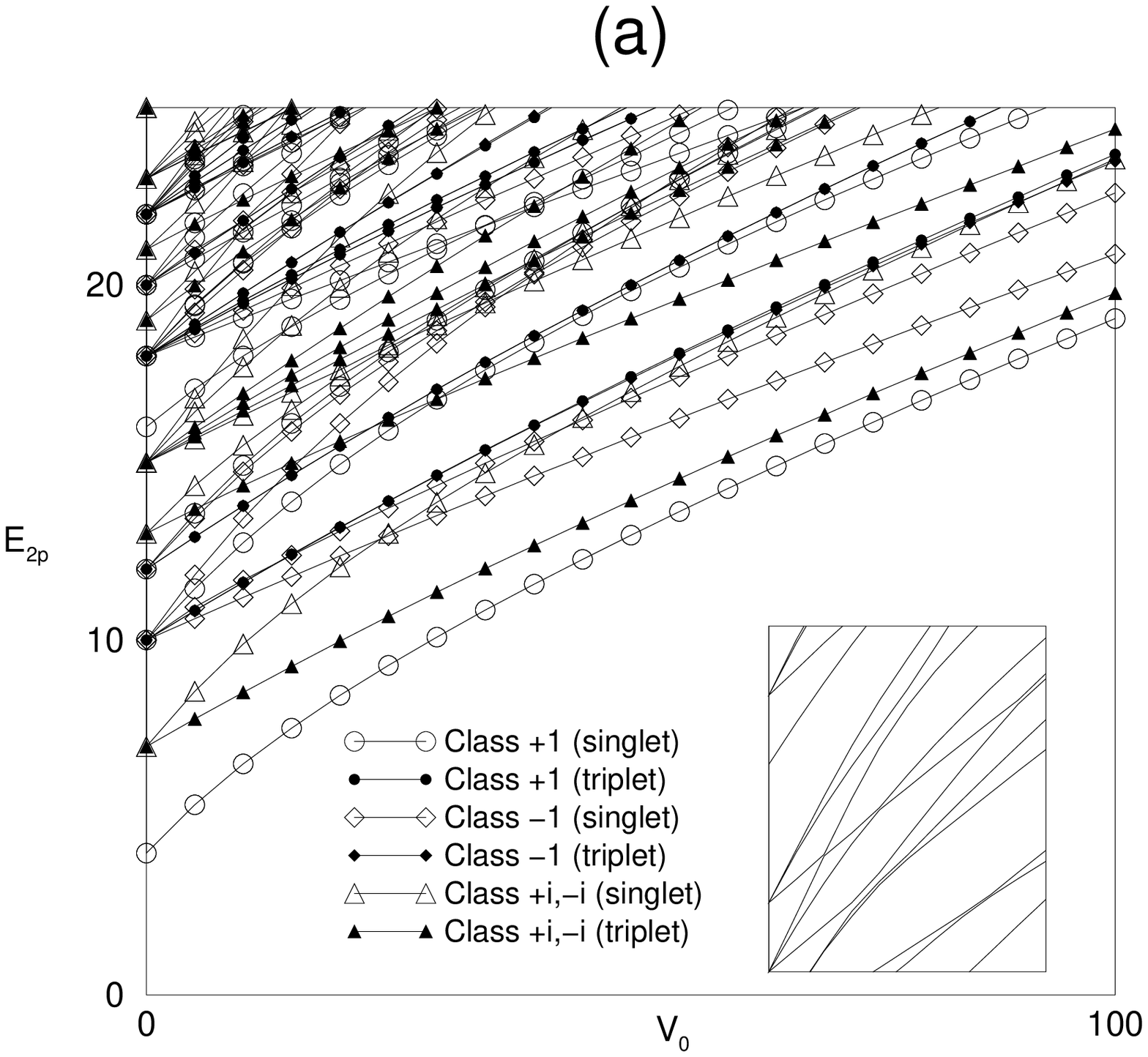}
\includegraphics[height=5cm,width=5cm]{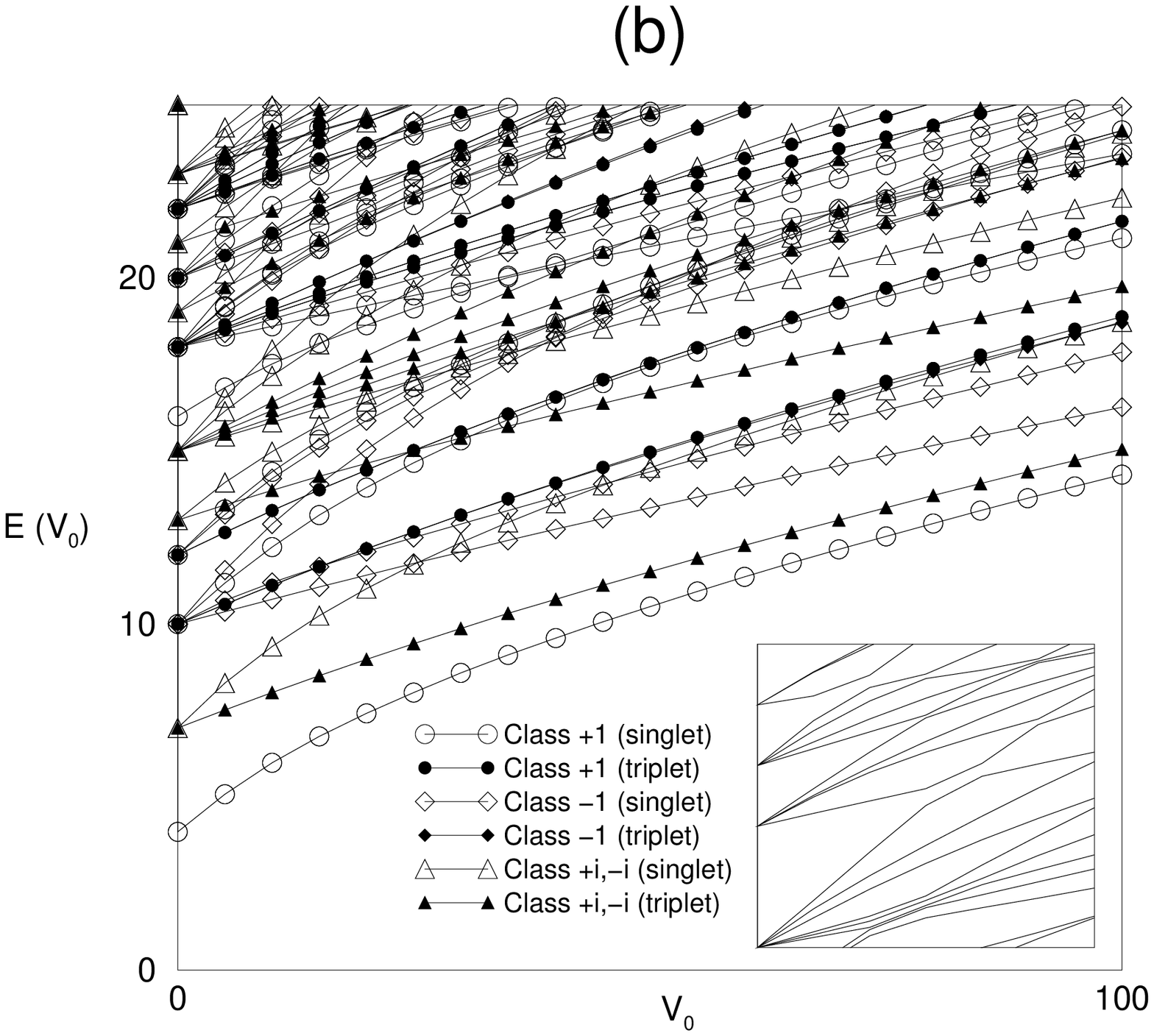}
\includegraphics[height=5cm,width=5cm]{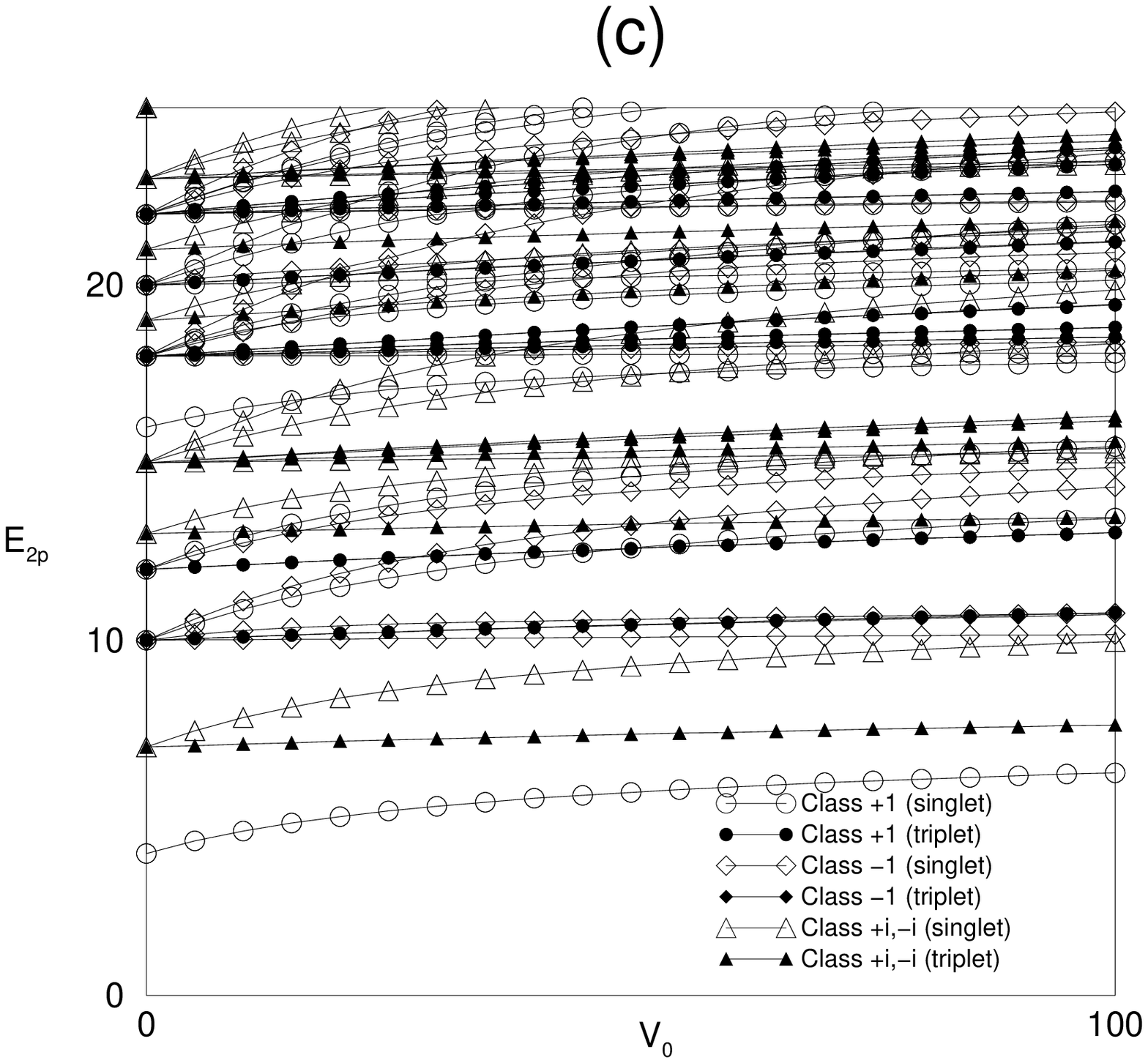}
\caption{ Energy levels as a function of $V_0$ for different values of
the reach parameter $\alpha$. (a) $\alpha=0$ (Coulomb interaction),
(b) $\alpha L=1$ and (d) $\alpha L=10$. Inset: avoided crossings on
the $(+1)$-singlet (solid line) and $(-1)$-singlet (filled squares)
classes}
\label{AE_V01}
\end{figure*}

Figure \ref{AE_V01} shows the two-particle energy levels as a function
of the interaction strength for different values of $\alpha L$. All
energy levels are shown, for the four symmetry classes of both singlet
and triplet configurations. The first panel shows the Coulomb case,
$\alpha=0$. The interaction induces singlet-triplet gaps
\cite{creffield2} and removes several degeneracies in the energy
levels. It also promotes level repulsion (``avoided crossings'')
within each symmetry class.  These are typical of systems with
GOE-type level spacing distribution. Although the number of levels
does not allow for a precise statistical analysis of the spectrum, the
level-spacing histograms (not shown) display a distinctive difference
between the non-interacting (Poisson-like) and the interacting
(GOE-like) cases. As the relative range decreases ($\alpha L$
increases) the levels become less sensitive to $V_0$ and the avoided
crossings narrow.

\subsection{Ground-state properties}
\label{gstate}

Recent works have investigated the formation of ``Wigner molecule"
type of ground states in polygonal quantum-dots in the low density
limit \cite{creffield1,grigorenko,akbar}. In this limit, the Coulomb
interaction between the electrons dominates over the kinetic energy
(the so called ``large $r_s$'' limit) and the ground state electron
density shows pronounced peaks near the corners of the polygonal
boundary \cite{creffield1}.

\begin{figure*}
\includegraphics[height=4cm]{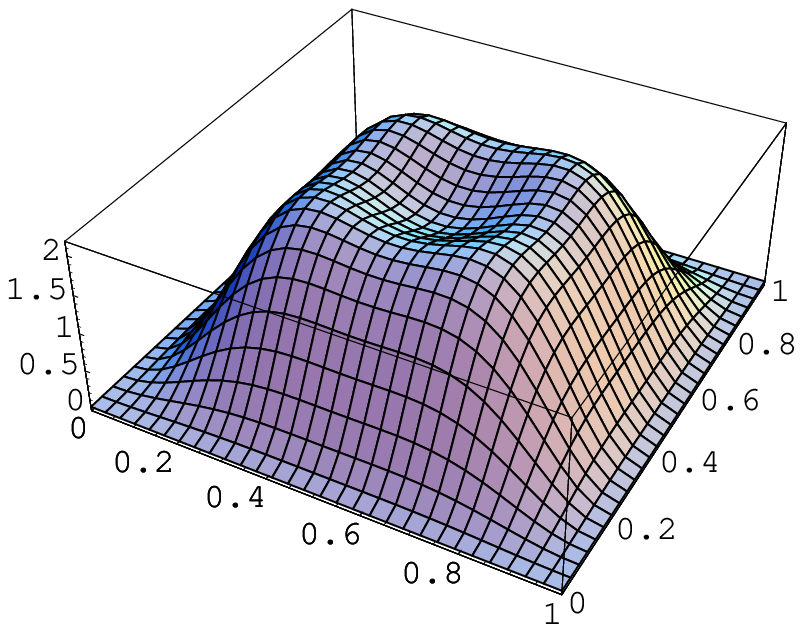}
\includegraphics[height=4cm]{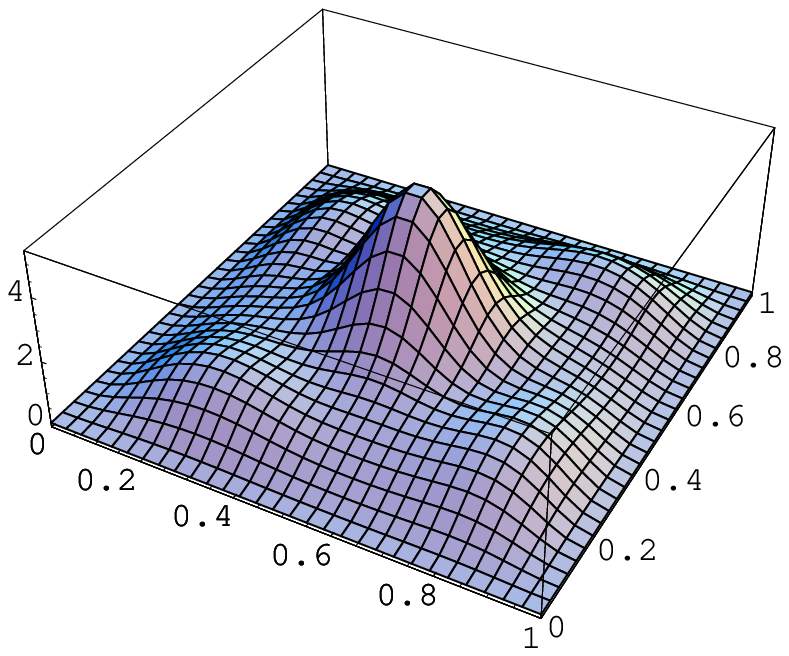}
\includegraphics[height=4cm]{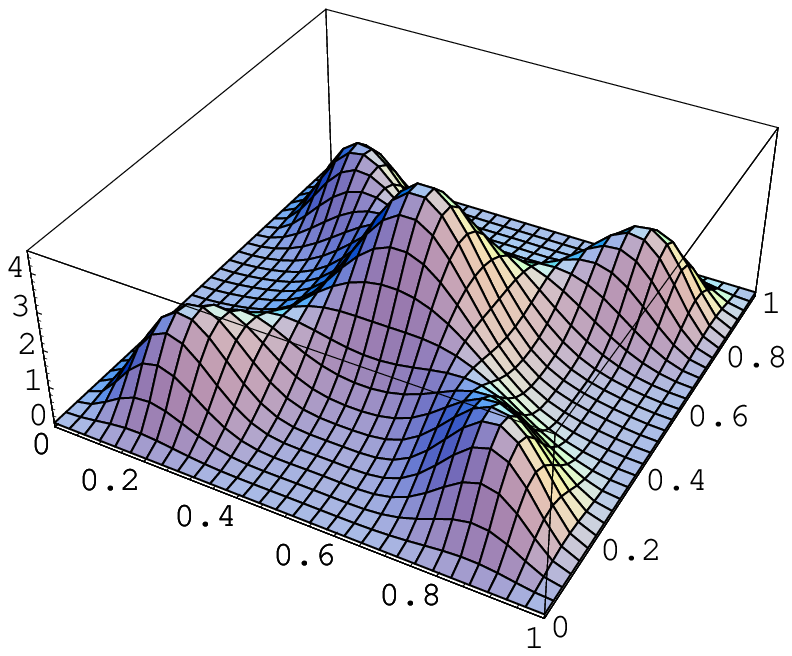}
\includegraphics[height=4cm]{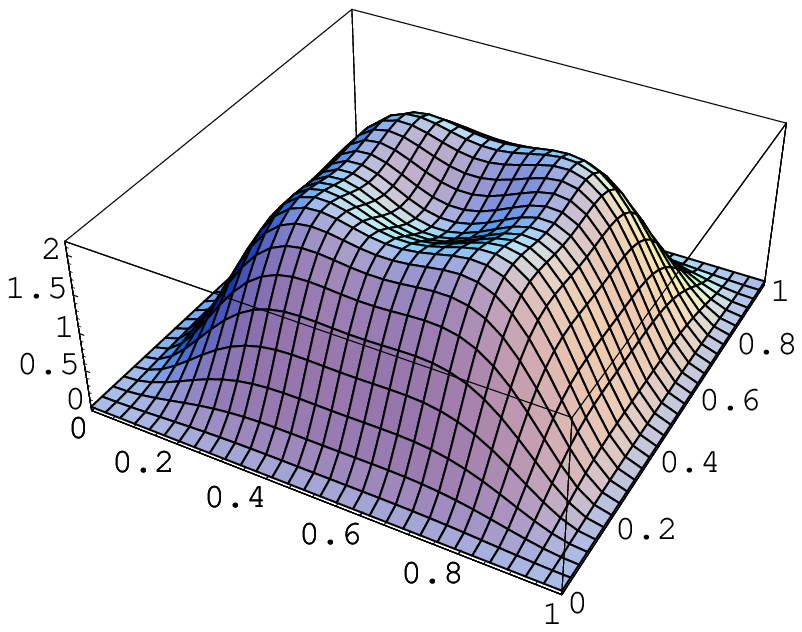}
\includegraphics[height=4cm]{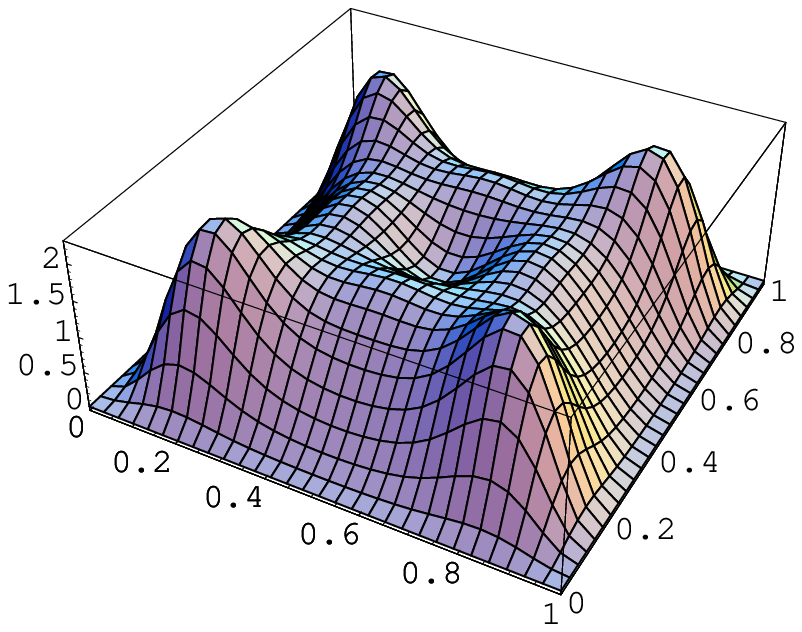}
\includegraphics[height=4cm]{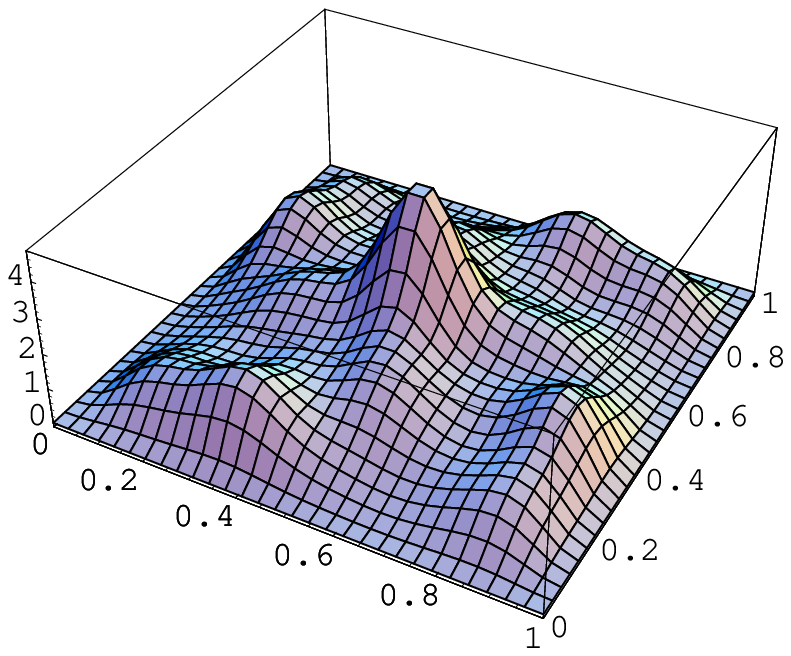}
\includegraphics[height=4cm]{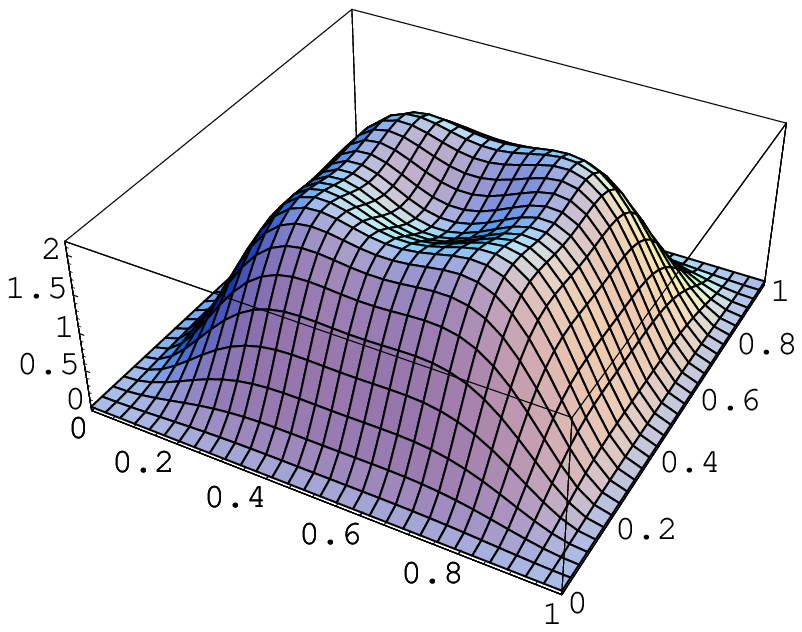}
\includegraphics[height=4cm]{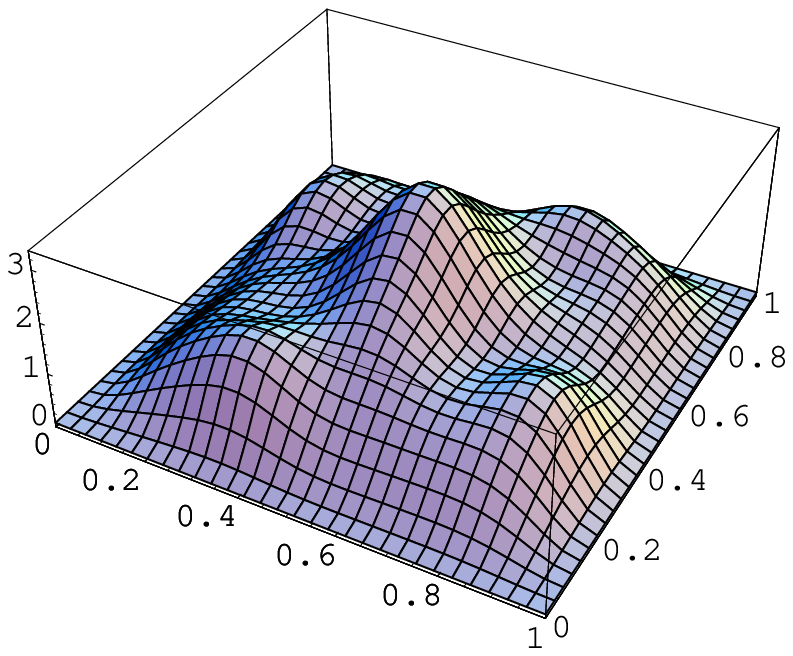}
\includegraphics[height=4cm]{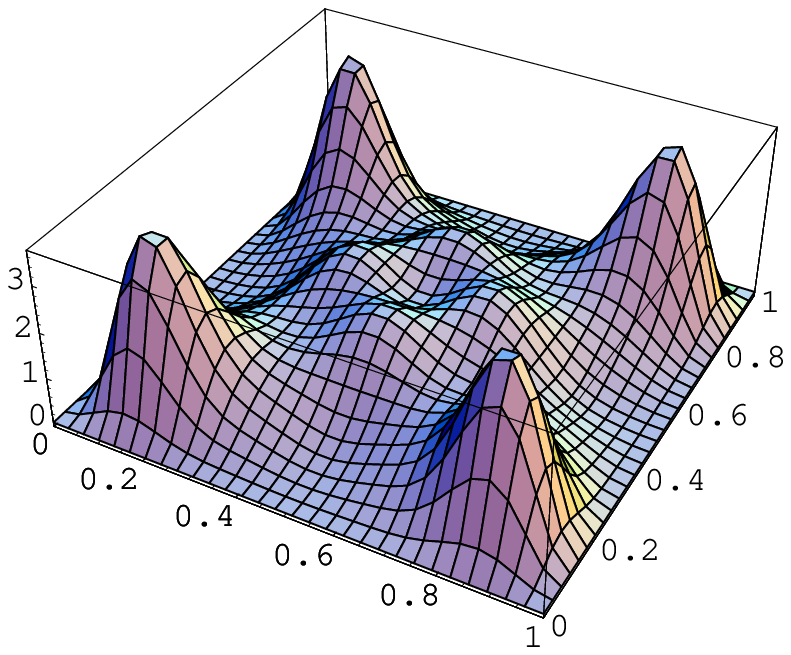}
\caption{ Electronic density for different values of $\alpha^{-1}$ and
$L$. For each line, the reach parameter $\alpha^{-1}$ is fixed for
increasing values of $L$ (from left to right $L=10,100$ and
$1000$nm). From top to bottom, we have $\alpha^{-1}=10,100$ and
$1000$nm. The Wigner Molecule state is recovered for $\alpha^{-1}=L=1
\mu m$}
\label{wig}
\end{figure*}

We have addressed the question of whether the finite-range character
of the repulsive potential would change such configuration. The
low-density limit can be approached by making $V_0 \rightarrow
\infty$. However, as discussed above, as the dot size $L$ increases, the
interaction strength increases but the effective interaction range
$\alpha^{-1}$ {\it decreases}. Figure \ref{wig} shows that, depending
on the value of $\alpha^{-1}$, the Wigner molecule state can be
suppressed, even for large dots. This figure shows the ground-state
electronic density
\begin{equation}
\rho(\vec{r_1}) =
  \int |\Psi_0 (\vec{r_1}, \vec{r_2})|^2 d \vec{r_2}
\label{dens}
\end{equation}
for $L=10$, $100$ and $1000$ $nm$ and $\alpha^{-1}=10$, $100$ and
$1000$ $nm$. Each column represents dots with the same width $L$ but
with different values of $\alpha$. Each line has the same value
of $\alpha$ but different sizes. The product $\alpha L$ is constant
along the diagonal. Even for the largest dot, with $1000 nm$, the
Wigner molecule state is completely suppressed for $\alpha L < 10$
(first two plots on the last column). Only when $\alpha L = 1$ (last
plot on the last column) does the electron density show pronunced
peaks near the corners.

\begin{figure*}
\includegraphics[height=4cm]{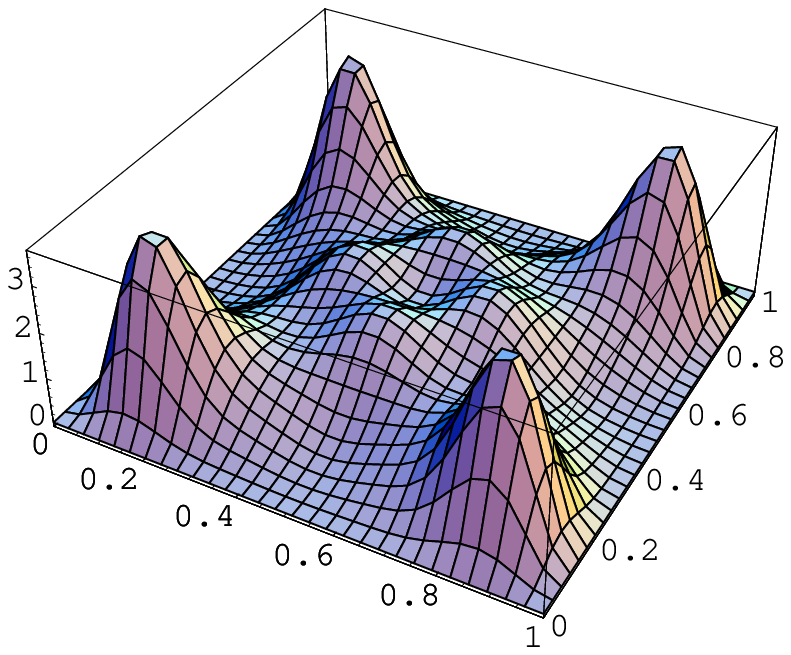}
\includegraphics[height=4cm]{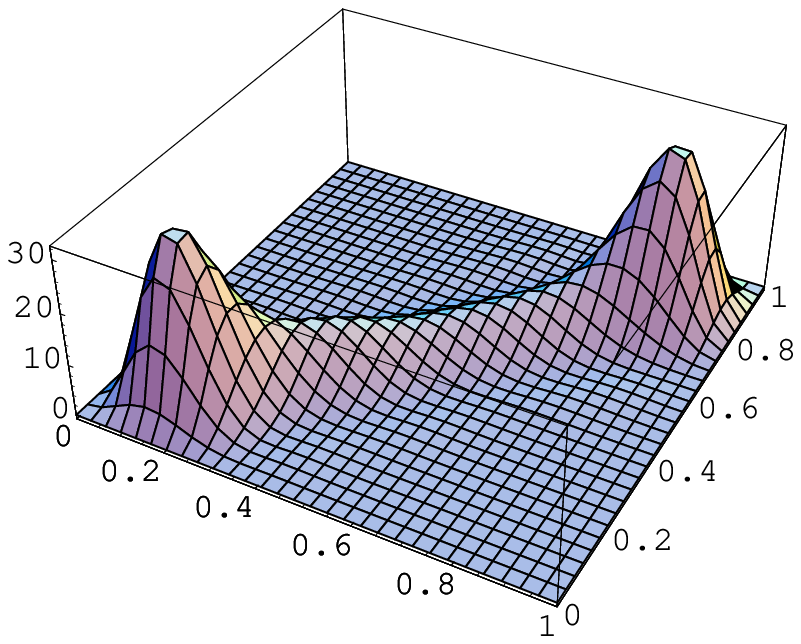}
\includegraphics[height=2.3cm]{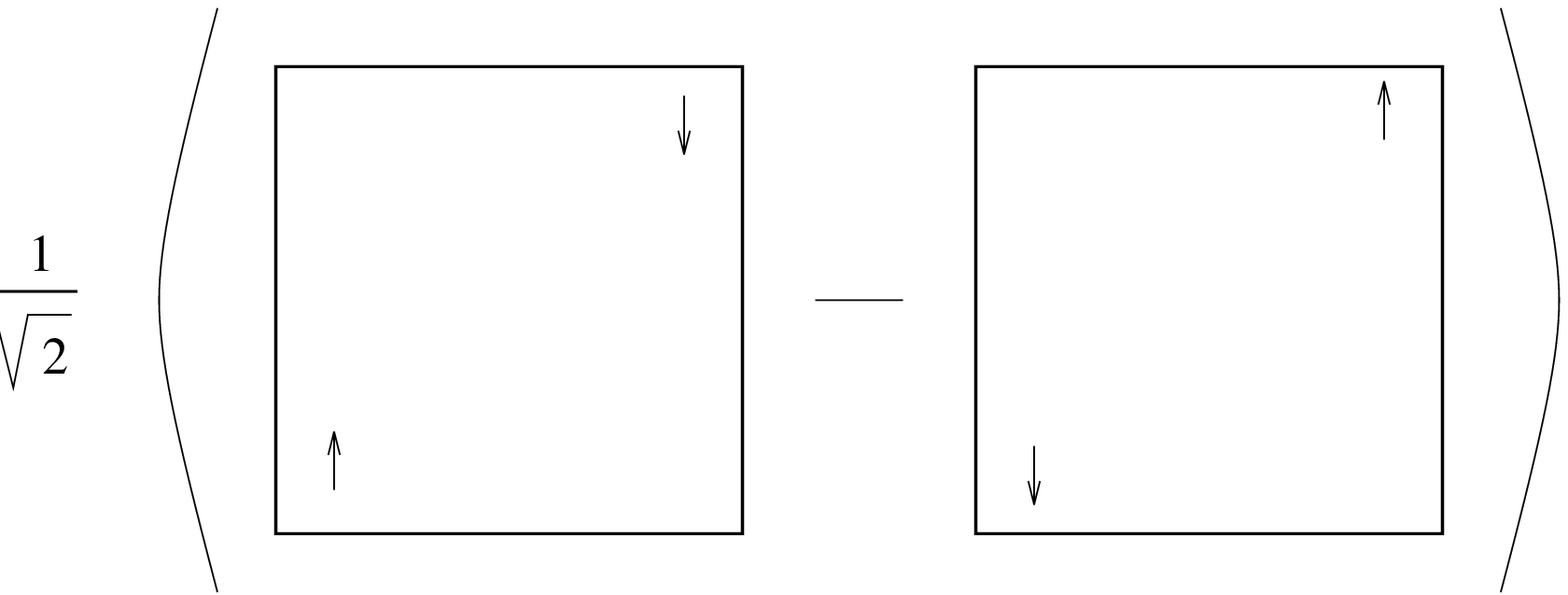} \\
\includegraphics[height=4cm]{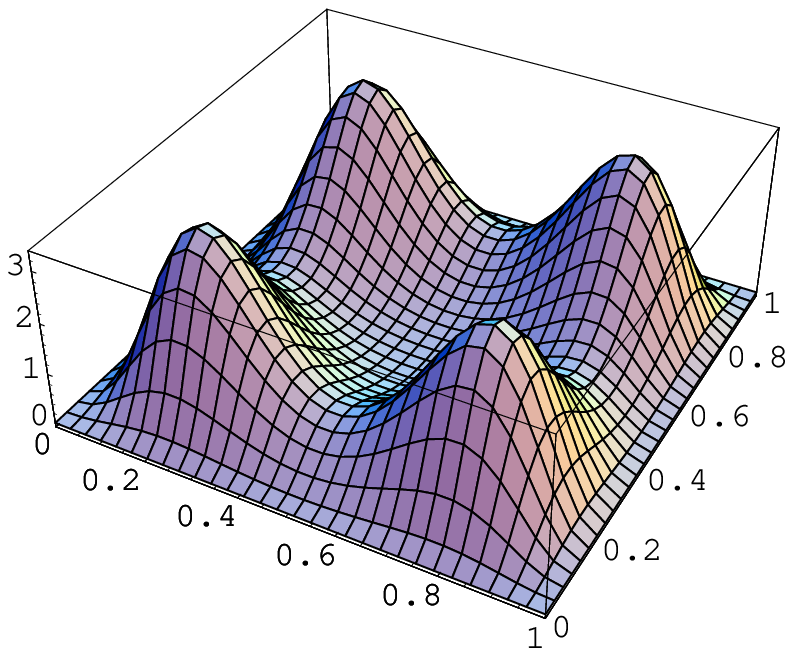}
\includegraphics[height=4cm]{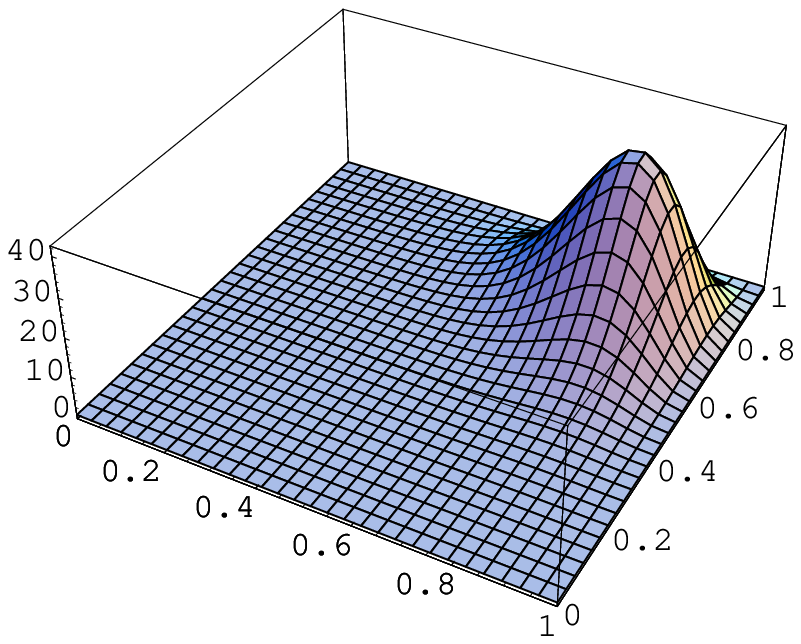} \qquad \qquad \qquad
\includegraphics[height=2.3cm]{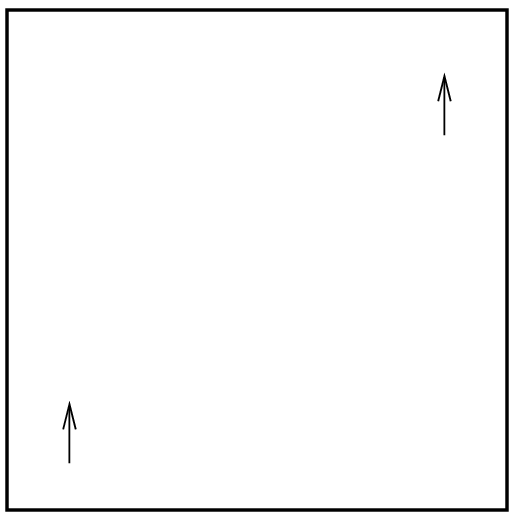} \qquad \qquad \qquad
\caption{ Top: Electronic density (right) and probability density with
one of the coordinates on a Wigner molecule peak ($|\Psi^{(C)}_0 ({\bf
r_1} = (0.2,0.2),{\bf r_2})|^2$) (middle) showing a singlet-like
spatial correlation (right). Bottom: Same for an excited triplet state
[(+i)-Class]}
\label{singlet}
\end{figure*}

Figure \ref{singlet} shows two examples of states with peaks near the
corners in the strong interaction limit. We see that, although the
electronic density of the two states looks similar, the spatial
correlations are very different, reflecting the fact that one of them
is a singlet and the other a triplet. Fixing ${\bf r}_1$ at the center
of one of the peaks, $\bar{{\bf r}}_1=(0.2,0.2)$, the probability
density $|\Psi ({\bf r_1} = \bar{{\bf r}}_1,{\bf r_2})|^2$ shows two
peaks along the diagonal for the singlet state and only one peak on
the opposite corner for the triplet state, since $\psi=0$ for ${\bf
r}_1={\bf r}_2$ in this case. The two-particle configuration is shown
schematically for both cases.


\section{Effects of the interaction in the magnetic properties}

\subsection{Energy levels}

\begin{figure*}
\includegraphics[height=6cm,width=5cm]{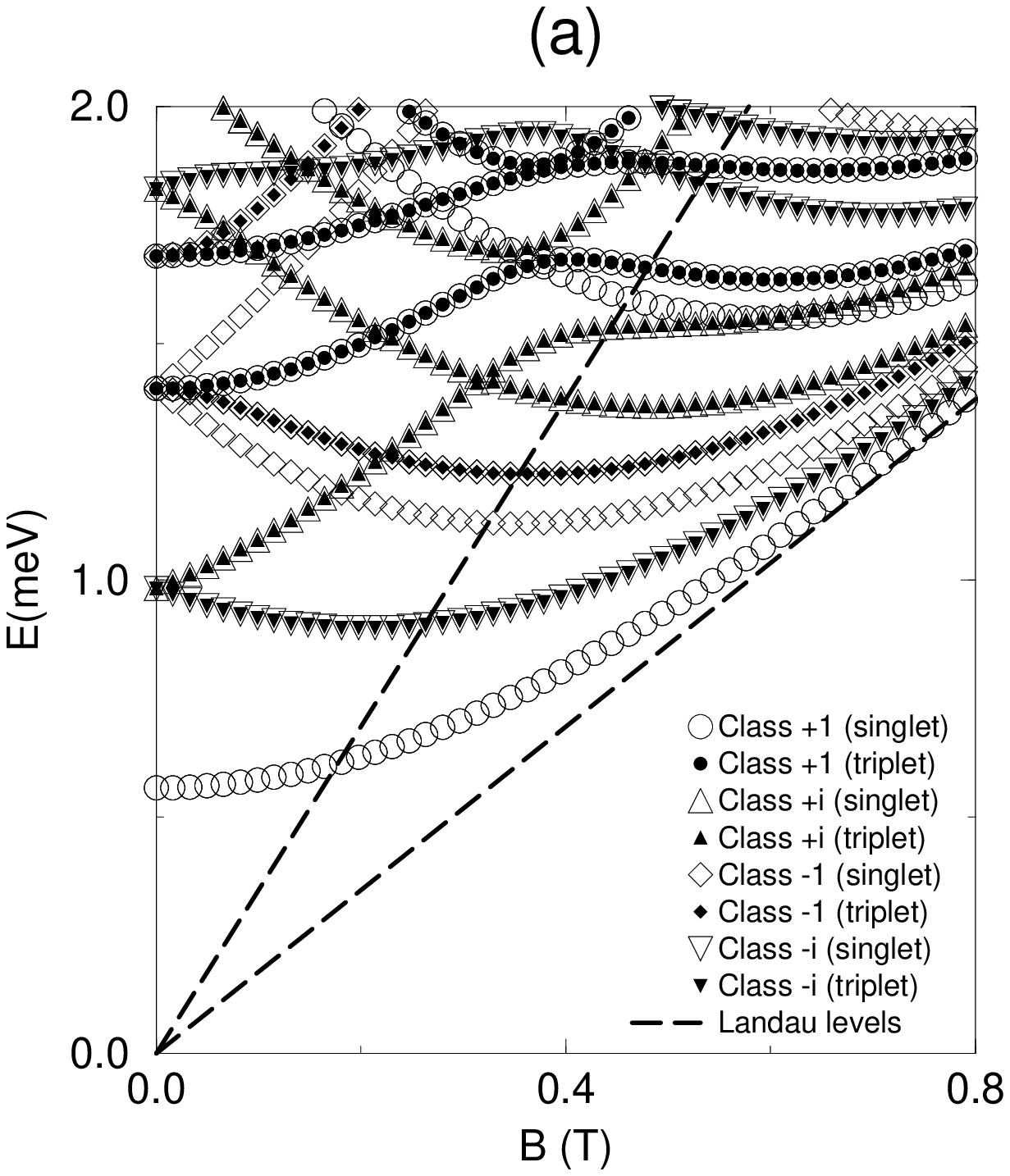}
\includegraphics[height=6cm,width=5cm]{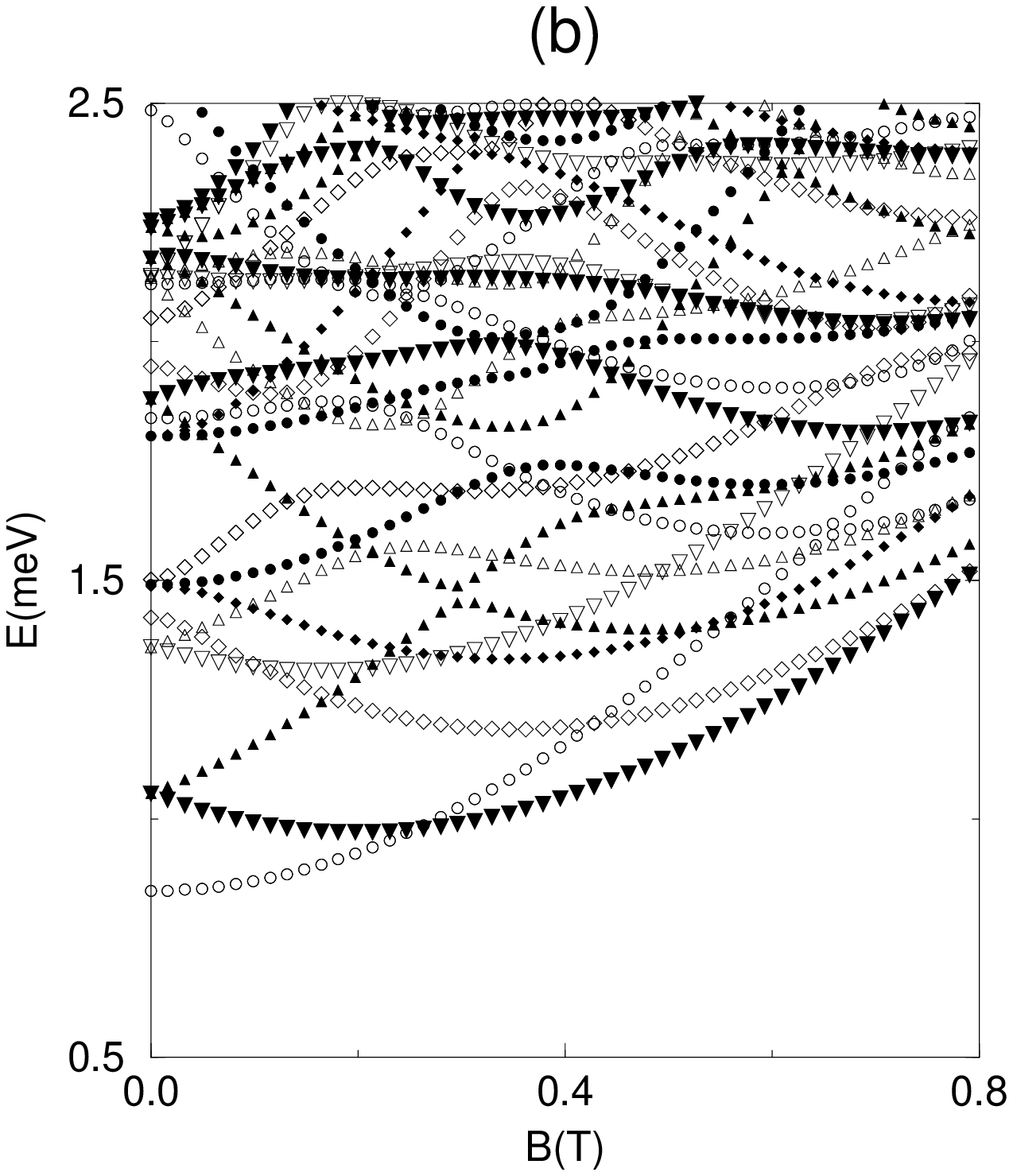}
\includegraphics[height=6cm,width=5cm]{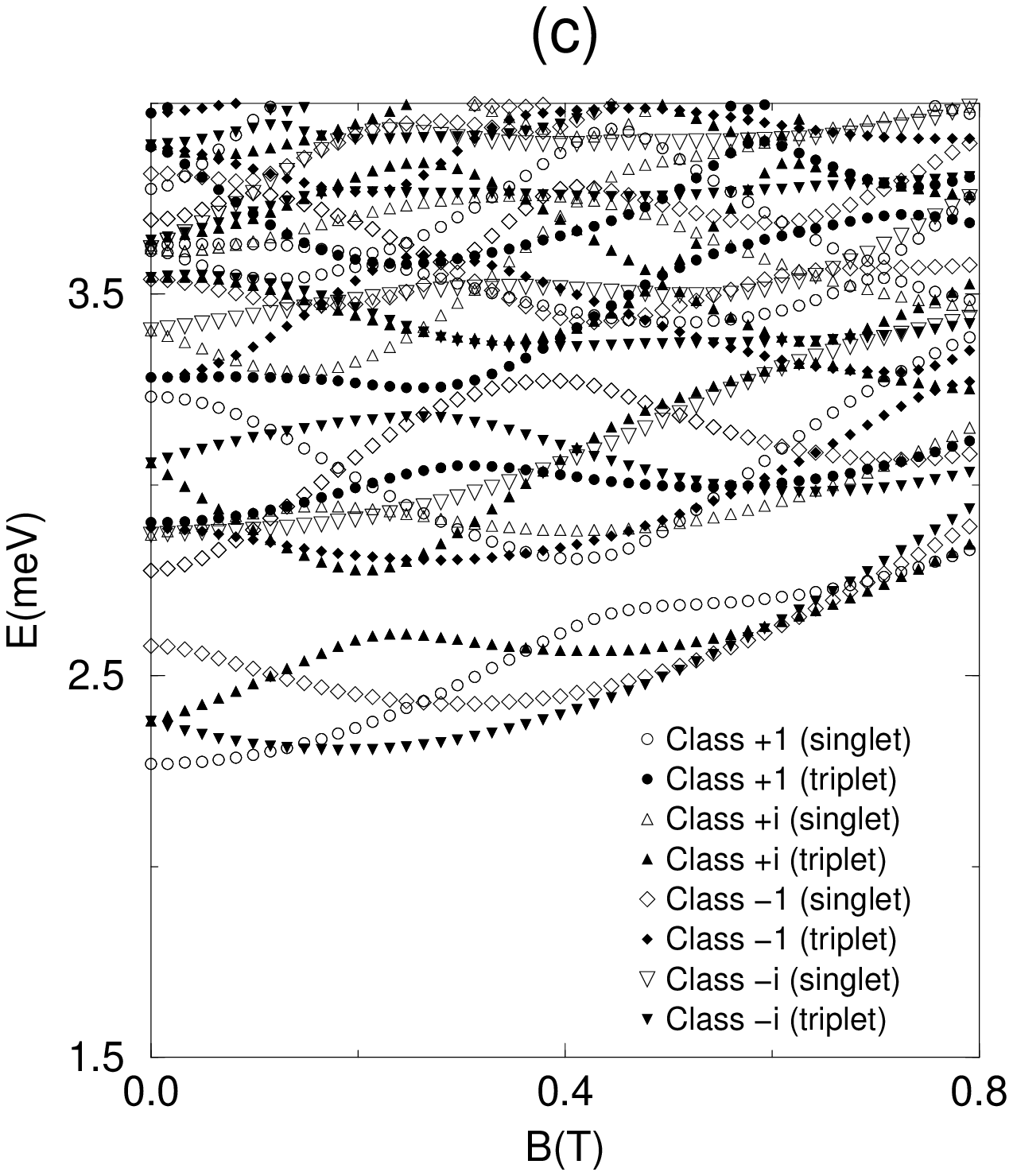}
\caption{ Energy levels as a function of $B$ for $L=200$nm and
different values of the reach parameter $\alpha^{-1}$. (a)
non-interacting ($V_0=0$), (b) $\alpha^{-1}=L/10$
and (c) Coulomb potential ($\alpha=0$)}
\label{E_BL200}
\end{figure*}

Figure \ref{E_BL200} shows the first energy levels as a function of the magnetic field for $L=200$
nm and different values of $\alpha$. The first plot shows the non-interacting case ($V_0=0$), where
the singlet and triplet two-particle levels are degenerate. Notice that the symmetry classes $(+i)$
and $(-i)$ are no longer degenerate for $B \neq 0$. Also, when the electronic interaction is
switched on (Figs.\ref{E_BL200}(b) and (c)), the singlet-triplet degeneracy is broken. The
combination of these two effects leads to singlet-triplet crossings in the ground state for
magnetic fields of the order of a few Tesla. This kind of oscillation has been studied previously
both theoretically (for the Coulomb case) \cite{wagner,peeters,ugajin,creffield2} and
experimentally \cite{ashoori,ashoori2}.

The role of the potential range $\alpha$ can also be seen from these figures. In
Fig.\ref{E_BL200}(b), where the range is only one tenth of the dot size, the splitting between the
$(+i)$ and $(-i)$ classes is still very clear, but the scale of the energy levels is much closer to
the non-interacting case. Also, and more importantly, there is only one singlet-triplet crossing in
the ground state, as opposed to the three crossings of the Coulomb case. For smaller values of
$\alpha^{-1}$ these crossings are completely suppressed.


\subsection{Partition function and susceptibility}

In this subsection we consider the orbital magnetization and magnetic
susceptibility of the interacting two-electron system. The partition
function $Z(B,T)=\mathit{Tr}\{e^{-\beta \hat{H}(B)}\}$ ($\beta \equiv
1/k_B T$) can be computed from the energy levels. The magnetization
$m(B)$ and the magnetic susceptibility $\chi(B)=\partial m(B)/\partial
B$ can be calculated from
\begin{equation}
m(B) = -\frac{1}{A} \; \frac{\partial F}{\partial B} =
\frac{1}{\beta A} \frac{\partial \log Z(B,T) }{\partial B} \;,
\end{equation}
where $F$ is the Helmholtz free energy and $A=L^2$.

\begin{figure*}
\includegraphics[height=6cm,width=5cm]{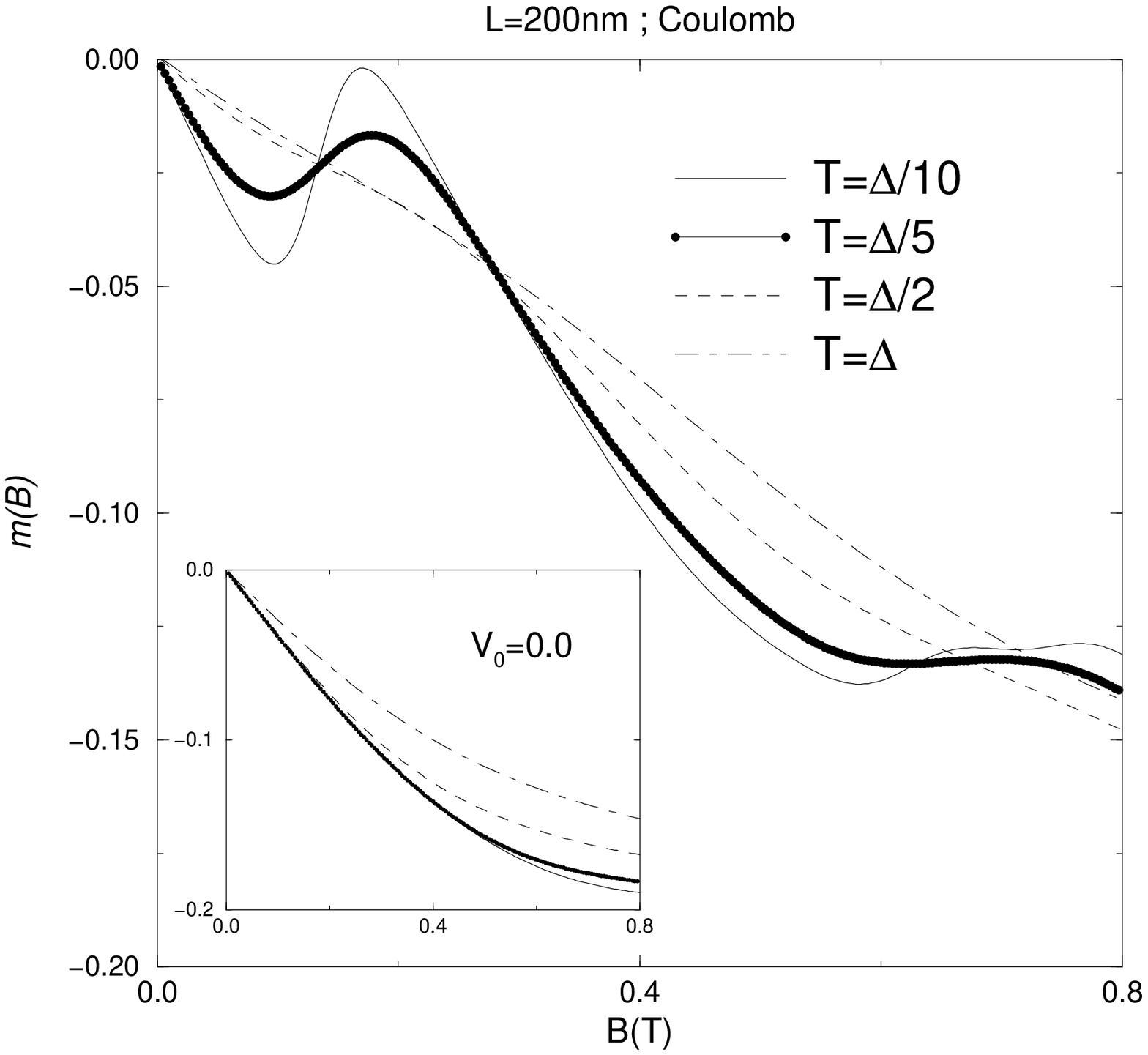}
\includegraphics[height=6cm,width=5cm]{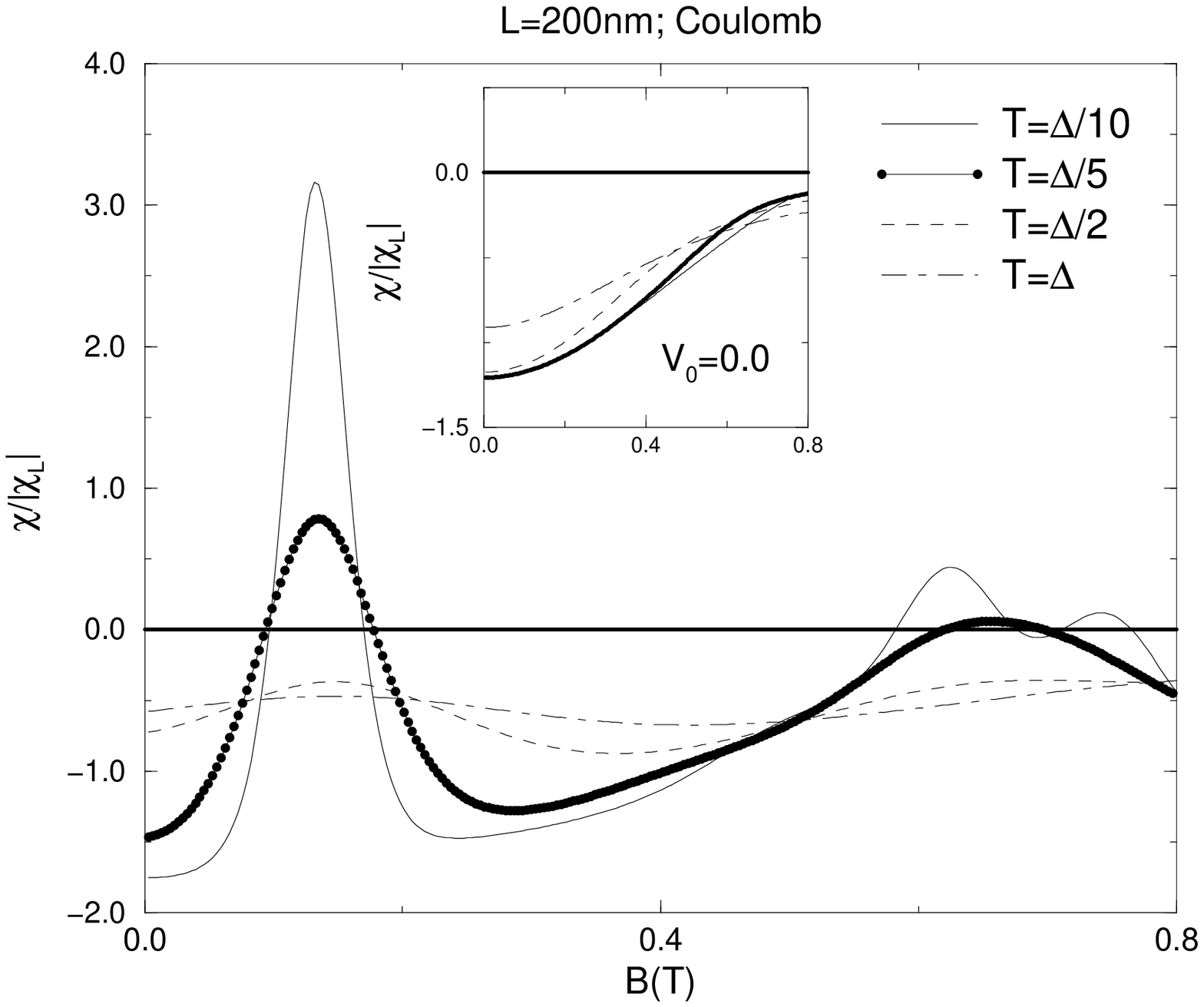}
\includegraphics[height=6cm,width=5cm]{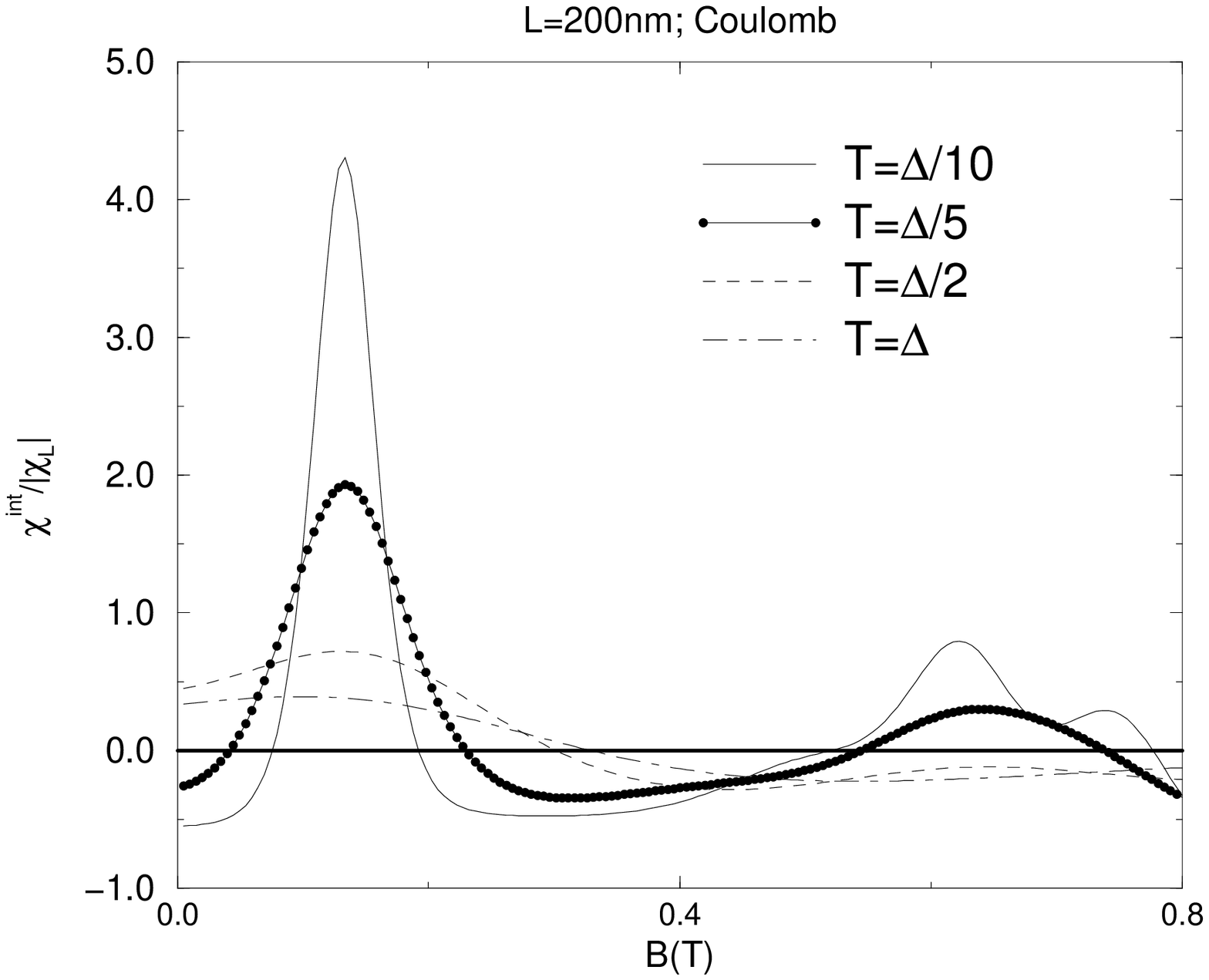} \\
\includegraphics[height=6cm,width=5cm]{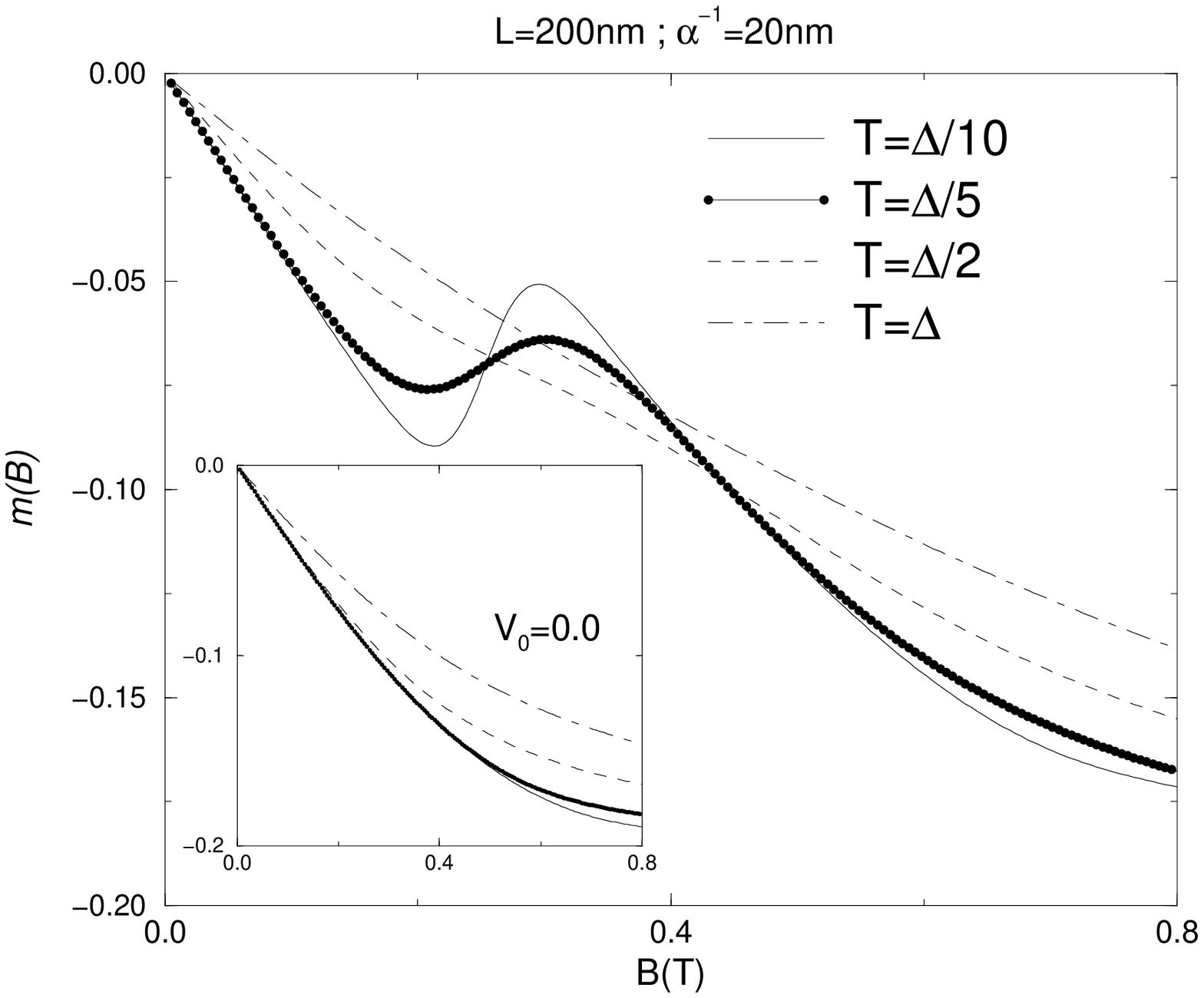}
\includegraphics[height=6cm,width=5cm]{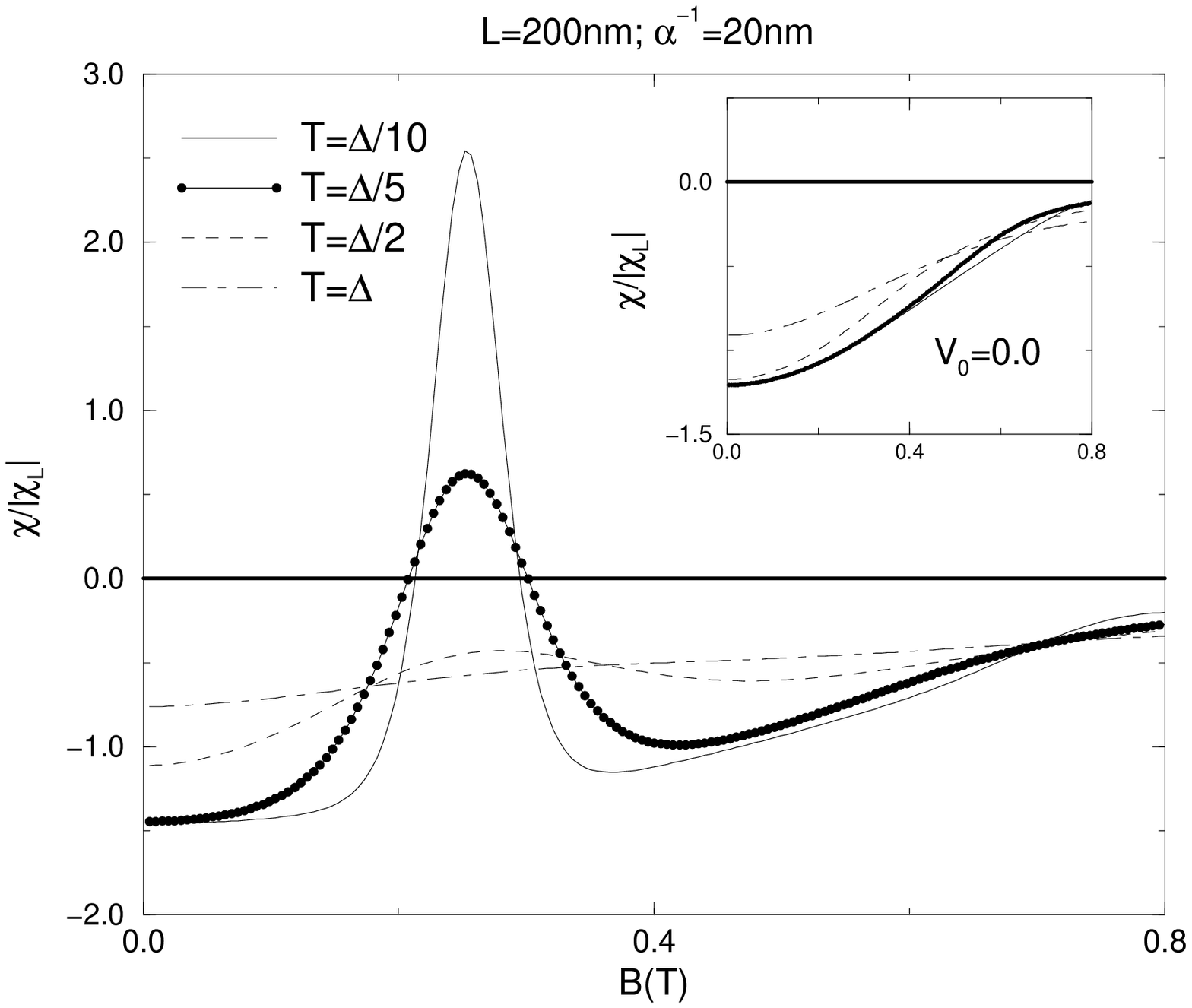}
\includegraphics[height=6cm,width=5cm]{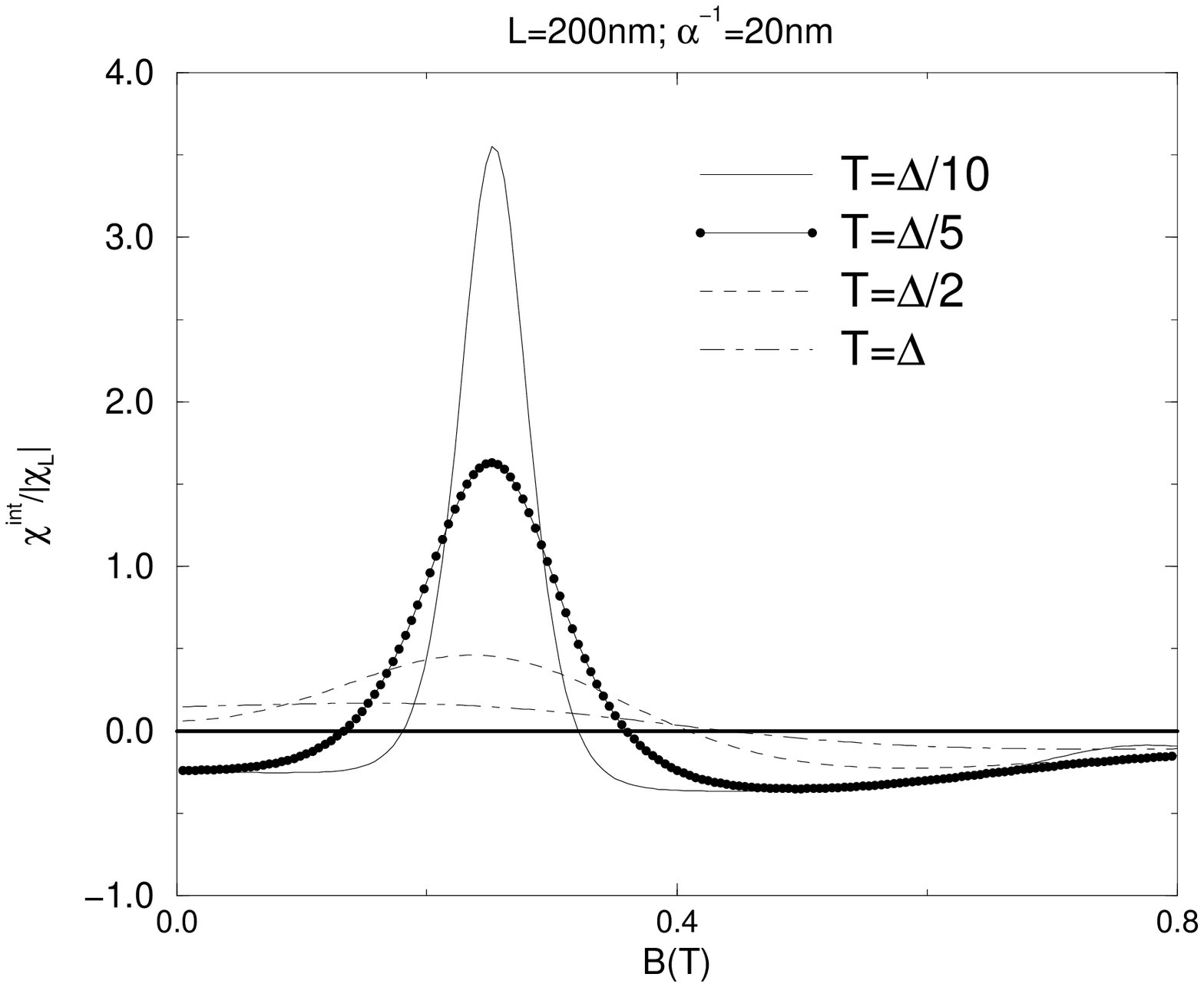}
\caption{ Magnetization $m(B)$, magnetic susceptibility $\chi(B)$ and
interaction induced susceptibility
$\chi^{\mbox{\scriptsize int}}=\chi-\chi^{(0)}$ as a function on the
magnetic field $B$ for a dot size of $L=200$nm and different values of
the potential reach parameter $\alpha$ (Top: $\alpha=0$ [Coulomb],
Bottom: $\alpha^{-1}=20$nm). For low temperatures $(T < \Delta/2)$,
the system features interaction-induced paramagnetic peaks. Inset: The
non-interacting magnetization and susceptibility.}
\label{mag_BL200}
\end{figure*}

Figure \ref{mag_BL200} shows the magnetization $m(B,T)$, the susceptibility $\chi(B,T)$ and the
interaction-induced susceptibility $\chi^{int}=\chi - \chi^{(0)}$ as a function of $B$ for
$\alpha=0$ (Coulomb potential) and $\alpha^{-1}=L/10$ (short-range interaction). $\chi^{(0)}$ is
the susceptibility of the non-interacting system. Both $\chi(B,T)$ and $\chi^{int}(B,T)$ are
expressed in units of the Landau susceptibility $|\chi_L|=e^2/12 \pi m^{*} c^2$ and the temperature
is expressed in units of the mean level spacing $\Delta$.

On the average, $\chi(B,T)$ is diamagnetic, as in the non-interacting
case (see $\chi^{(0)}$ on the inset). However, the exchange-induced
singlet-triplet crossings of the ground state level contributes
paramagnetic fluctuations of order of $\sim 3 \chi_L$ at low
temperatures. As discussed in the previous subsection, these crossings
tend to disappear for short range interactions. For $\alpha L=10$ only
one crossing survives.

For very low temperatures, only the ground state contributes
significantly to the partition function and $F \approx E_0$. In the
region close to the crossing, $\partial E_0/\partial B$ is
discontinuous with a negative curvature, which explains the
paramagnetic peak in $\chi(B)$. For higher temperatures, the
``anti-crossing'' (positive curvature) of the first excited state
tends to compensate the ground state crossing and the fluctuations are
attenuated.

\begin{figure*}
\includegraphics[height=10cm]{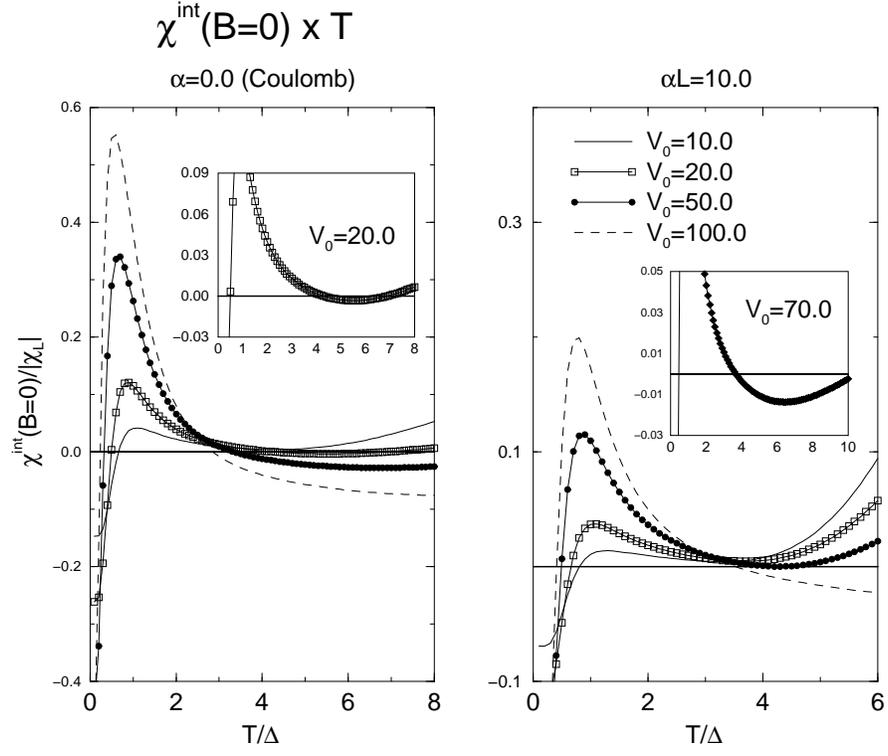}
\caption{ Interaction-induced zero field susceptibility
$\chi^{\mbox{\scriptsize int}}_{B=0}(T)$ as a function of the
temperature for a Coulomb potential (left) and a Yukawa potential with
$\alpha^{-1}=L/10$ (dir.) and different values of the potential
strength $V_0$. }
\label{Dsus0_T}
\end{figure*}

Figure \ref{Dsus0_T} shows the behavior of the interaction-induced
susceptibility at zero magnetic field $\chi^{int}(B=0,T)$ for the
Coulomb the and short-range ($\alpha^{-1}=L/10$) potentials and
different values of the strength parameter $V_0$. The results show
paramagnetic and diamagnetic phases as a $T$ increases, with a
diamagnetic minimum at $T \approx 2 \Delta$. For low values of $V_0$
the susceptibility is again paramagnetic for high $T$. This type of
behavior has been observed before for weak contact (Dirac delta) type
of interactions \cite{harold}. Our results show that the existence of
paramagnetic and diamagnetic phases in $\chi^{int}(B=0,T)$ might be
more general, and not so sensitive to the type of interaction.

\section{Discussion}

This work was motivated in part by the experimental results of Levy et
al \cite{levy}, who measured the magnetic properties of an ensemble of
square dots containing a few thousands electrons each in the balistic
regime. Our objective in this paper was to understand the effects of
the residual electronic interaction in the simplest possible case,
i.e., that of only two electrons. We simulated the shielding of the
bare Coulomb force by using an exponential type of cut-off, like that
of the Yukawa potential. The range of the interaction, $\alpha$, was
considered as a free parameter. The size of the dot, which controls
the relative strength of the potential, $V_0,$ with respect to the
kinetic energy, acts as a second parameter. Our results show that both
$V_0$ and $\alpha$ are important to determine the properties of the
energy spectrum and of the probability profile of the ground state. We
showed, in particular, that short range interactions may suppress the
appearance of Wigner molecule type of states even for strong
interactions ($V_0$ large).  The range of the interaction also affects
the magnetic susceptibility of the system. Short range interactions
might inhibit singlet-triplet oscillations of the ground state,
suppressing the paramagnetic fluctuations of $\chi(B)$. Finally we
have shown that the part of the susceptibility induced by the
interaction presents paramagnetic and diamagnetic phases as a function
of the temperature, in agreement with the results obtained by a
pure Dirac delta interaction \cite{harold}.

It is difficult speculate at this point if our results point to an
explanation of the slow decay of the susceptibility with the
temperature, as observed experimentally by Levy et al. In a system with
many electrons, the ground state energy oscillates as a function of
$B$ even without electronic interaction. This type of oscillation is
only due to the boundary, and is responsible for the paramagnetic
susceptibility of the gas. Our results indicate that, besides this
``boundary-induced'' effects, the electronic interaction is
responsible for further oscillations, this time between singlet and
triplet states. These ``interaction-induced'' oscillations also
contribute to the susceptibility. In the present case of two electrons
this contribution turns out to be larger than that of the
non-interacting case. Besides, the curves in Fig. \ref{Dsus0_T} show
that $\chi^{int}(B=0,T)$ is sensitive to both $V_0$ and $\alpha$. The
peak of $\chi^{int}(B=0,T)$ at $T \approx \Delta$ leads to a slightly
slower decay of the overall susceptibility. Although this result
might be peculiar of few-particle systems, there are similarities
between our findings and those obtained from semiclassical analysis
and RPA perturbation theory in the high-density limit \cite{harold},
such as the diamagnetic minima in $\chi^{int}(B=0,T)$. This might
indicate that the interaction indeed plays an important role in
behavior of the susceptibility with the temperature, although
calculations with more electrons should be conducted to confirm this
conjecture.


\centerline{Acknowledgements}
\noindent This paper was partly supported by {\bf FAPESP}, {\bf CNPq} and {\bf Finep}. LGGVDS would
like to thank Prof. Harold U. Baranger for his contributions and support to the development of this
work and acknowledges the hospitality of the Department of Physics at Duke University. We also
would like to thank Dennis Ullmo, Gonzalo Usaj and Charles Creffield for helpful discussions and
suggestions.

\newpage
\appendix

\section{Calculation of the integrals}

In order to calculate the matrix element (\ref{Integral}) we first
notice that the integrand can be decoupled using relative and
center-of-mass coordinates. Since the masses are equal (we set
$m_1=m_2 \equiv 1$ for simplicity), we have $\vec{r} = \vec{r_2} -
\vec{r_1}$ and $\vec{R} = (\vec{R_2} + \vec{R_1})/2$.
When the sine functions are written as a sum of complex exponentials
\bd
\begin{array}{rcl}
\sin{2 m \pi x_i} & = & \sum_{\epsilon_i = -1}^{1}
\frac{\epsilon_i \; e^{\epsilon_i i M_i x_i}}{2i}\\[0.2truecm]
\sin{2 n \pi y_i} & = & \sum_{\eta_i = -1}^{1} \frac{\eta_i \;
e^{\eta_i i N_i y_i}}{2i} \;,
\end{array}
\ed
the integrand in Eq.(\ref{Integral}) becomes
\begin{eqnarray}
\lefteqn{ \sum_{\epsilon_i \eta_i} \frac{(\epsilon_1 \cdots \epsilon_2^{'}) (\eta_1 \cdots
\eta_2^{'})}{256} \times} \nonumber \\
&& \left(  e^{i(\alpha_1+\alpha_2)X}  e^{i(\beta_1+\beta_2)Y} e^{i(\alpha_2-\alpha_1)\frac{x}{2}}
e^{i(\beta_2-\beta_1) {\frac{y}{2}}} \frac{e^{-\alpha r}}{r} \right) \nonumber \\
  \label{}
\end{eqnarray}
where
\begin{equation}
\begin{array}{l}
\alpha_1 =  \frac{\pi m_1}{L} \epsilon_1 +
\frac{\pi m_1^{'}}{L}  \epsilon_1^{'} \;, \qquad
\alpha_2 = \frac{\pi m_2}{L} \epsilon_2 +
\frac{\pi m_2^{'}}{L}  \epsilon_2^{'} \\
\beta_1  =   \frac{\pi n_1}{L} \eta_1 + \frac{\pi n_1^{'}}{L}
\eta_1^{'} \; , \qquad
\beta_2  =   \frac{\pi n_2}{L} \eta_1 + \frac{\pi n_2^{'}}{L}
\eta_2^{'} \\
\end{array}
\label{alphas_betas}
\end{equation}

The integrals on the Center-of-Mass coordinates can then be readly
evaluated, at the cost of working on the complex plane. However, one
should note that the limits of integration of $(x,y)$ and $(X,Y)$ are
not independent. In the rest of this development, we will take $L
\equiv 1$ for the sake of simplicity. This is equivalent to perform
the calculations on a adimensional variable $\frac{r}{L}$. In order to
consider specific sizes for the dot, one has to include an additional
factor of $L$ multiplying the whole integral and scale $\alpha$
accordingly.

The change of variables $(x_1,y_1,x_2,y_2) \rightarrow (x,y,X,Y)$
leads us to four different sets of integration limits (the quadrants
on the $(x,y)$ plane). Next, we show the calculation of the integral
on the first quadrant $(I_1)$. The calculation on the other quadrants
($I_2,I_3,I_4$) is analogous. The total integral is then $I=I_1+I_2+I_3+I_4$.
On the first quadrant we have

\begin{eqnarray}
\lefteqn{I_1 = \sum_{\epsilon_i \eta_i} \frac{(\epsilon_1 \cdots \epsilon_2^{'}) (\eta_1 \cdots
\eta_2^{'})}{256} \times } \nonumber \\
&& \! \! \! \! \! \! \! \left( \int dA_1 e^{i(\alpha_1+\alpha_2)X}  e^{i(\beta_1+\beta_2)Y}
e^{i(\alpha_2-\alpha_1)\frac{x}{2}}  e^{i(\beta_2-\beta_1) {\frac{y}{2}}} \frac{e^{-\alpha
\sqrt{x^2+y^2}}}{\sqrt{x^2+y^2}} \right) \nonumber \\
\! \! \! \! \! \label{Ik}
\end{eqnarray}
where
\begin{equation}
\int dA_1 = \int_{0}^{1} dx \int_{0}^{1} dy
\int_{\frac{x}{2}}^{1-\frac{x}{2}} dX
\int_{\frac{y}{2}}^{1-\frac{y}{2}} dY \,.
\label{intgeral}
\end{equation}

After doing the integrals on the center-of-mass coordinates we are
left with
{\small
\begin{equation}
I_1 = \sum_{\epsilon_i \eta_i} \frac{(\epsilon_1 \cdots \epsilon_2^{'})
(\eta_1 \cdots \eta_2^{'})}{256}
\left\{ \int_{0}^{1} \int_{0}^{1} f(x,y)
\frac{e^{-\alpha \sqrt{x^2 + y^2}}}{\sqrt{x^2 + y^2}} \; dx dy \right\}
\label{intk2d}
\end{equation}
}
where the exact format of $f(x,y)$ depends on whether we have
$(\alpha_1 + \alpha_2)$ or $(\beta_1 + \beta_2)$ equal to zero or
not. We are going to write it down explicitly in a moment.

A transformation to relative polar coordinates yields:
{\small
\begin{equation}
I_1 = \sum_{\epsilon_i \eta_i} \frac{(\epsilon_1 \cdots \epsilon_2^{'})
(\eta_1 \cdots \eta_2^{'})}{256}
\left\{ \int_{0}^{\frac{\pi}{2}} \int_{0}^{r_1(\theta)}
f(r,\theta) e^{-\alpha r}  \;  dr d\theta \right\}
\end{equation}
}
where $r_1(\theta)=1/\cos{\theta}$ for $0< \theta < \frac{\pi}{4}$ and
$r_1(\theta)=1/\sin{\theta}$ for $\frac{\pi}{4} < \theta < \frac{\pi}{2}$.
The integral over $r$ can be done analytically and we are left with a
set of integrals over $\theta$:
\begin{equation}
I_1 = \sum_{\epsilon_i \eta_i} \frac{(\epsilon_1 \cdots \epsilon_2^{'})
(\eta_1 \cdots \eta_2^{'})}{256}
\left\{ \int_{0}^{\frac{\pi}{2}} F(\theta) d\theta \right\}
\label{intk2d_polar}
\end{equation}
As one can see from eq. (\ref{Ik}), the explicit form of $F(\theta)$
depends whether $(\alpha_1+\alpha_2)$ and/or $(\beta_1+\beta_2)$ are
equal to zero or not. We now present the specific form of $F(\theta)$
for all different possibilities.


\begin{widetext}
\subsection{$(\alpha_1+\alpha_2),(\beta_1+\beta_2) \neq 0$}
\begin{equation}
F(\theta)   =   \frac{(-1)}{(\alpha_1 + \alpha_2)(\beta_1 + \beta_2)} \left[  \frac{ e^{(\alpha_1 +
\alpha_2 + \beta_1 + \beta_2)} \left( e^{Z_a r_1 } - 1 \right)}{Z_a (\theta)} + \frac{\left( e^{Z_b
 r_1 } - 1 \right)}{Z_b (\theta)} + \frac{e^{(\alpha_1 + \alpha_2)} \left( 1 -
e^{Z_c  r_1 } \right)}{Z_c (\theta)} +  \frac{e^{(\beta_1 + \beta_2)} \left( 1 - e^{Z_d r_1 }
\right)}{Z_d (\theta)} \right]
\end{equation}
%
\subsection{$(\alpha_1+\alpha_2) = (\beta_1+\beta_2) =  0$}
\begin{equation}
 F(\theta)   =    \frac{\left( e^{z_a  r_1 } - 1 \right)} {z_a (\theta)} -
\frac{(\cos{\theta} + \sin{\theta})}{z_a^2 (\theta)} \left[  \left(e^{z_a  r_1} (z_a  r_1 - 1 ) +1
\right)\right]  + \frac{(\cos{\theta} \sin{\theta})}{z_a^3 (\theta)} \left[ \left(e^{z_a r_1}
(z_a^{2} r^{2}_1 - 2 z_a r_1 + 2 ) -2 \right) \right]
\end{equation}
\subsection{$(\alpha_1+\alpha_2) = 0$ and $(\beta_1+\beta_2) \neq  0$}
\begin{equation}
F(\theta) =  e^{(\beta_1 + \beta_2)} \frac{\left( e^{z_b  r_1 } - 1 \right)}{z_b (\theta)} -
\frac{\left( e^{z_c  r_1 } - 1 \right)}{z_c (\theta)} - \cos{\theta} \left[ e^{(\beta_1 + \beta_2)}
\frac{\left(e^{z_b  r_1 } (z_b  r_1  - 1 ) +1 \right)}{z_b^{2}} - \frac{\left(e^{z_c   r_1 } (z_c
 r_1  - 1 ) +1 \right)}{z_c^{2}} \right]
\end{equation}
\subsection{$(\alpha_1+\alpha_2) \neq 0$ and $(\beta_1+\beta_2) = 0$}
\begin{equation}
F(\theta) =   e^{(\alpha_1 + \alpha_2)} \frac{\left( e^{z_d r_1 } - 1 \right)}{z_d (\theta)} -
\frac{\left( e^{z_e r_1 } - 1 \right)}{z_e (\theta)} -  \sin{\theta} \left[ e^{(\alpha_1 +
\alpha_2)} \frac{\left(e^{z_d  r_1  } (z_d  r_1   - 1 ) +1 \right)}{z_d^{2}} - \frac{\left(e^{z_e
  r_1  } (z_e  r_1  - 1 ) +1 \right)}{z_e^{2}} \right]
\end{equation}
\end{widetext}
where
\[ \begin{array}{rcl} Z_a (\theta) & = & -\alpha -
i\left(\alpha_1\cos{\theta} + \beta_1\sin{\theta} \right) \\
Z_b (\theta) & = & -\alpha +
i\left(\alpha_2\cos{\theta} + \beta_2\sin{\theta} \right) \\
Z_c (\theta) & = & -\alpha -
i\left(\alpha_1\cos{\theta} - \beta_2\sin{\theta} \right) \\
Z_d (\theta) & = & -\alpha + i\left(\alpha_2\cos{\theta} - \beta_1\sin{\theta} \right) \\
\\
z_a(\theta) & \equiv & -\alpha + i(\alpha_2 \cos{\theta} +
\beta_2 \sin{\theta})\\
z_b(\theta) & \equiv & -\alpha + i(\alpha_2 \cos{\theta} -
\beta_1 \sin{\theta}) \\
z_c(\theta) & \equiv & -\alpha + i(\alpha_2 \cos{\theta} +
\beta_2 \sin{\theta}) \\
z_d(\theta) & \equiv & -\alpha + i(\beta_2 \sin{\theta} -
\alpha_1 \cos{\theta}) \\
z_e(\theta) & \equiv & -\alpha + i(\beta_2 \sin{\theta} +
\alpha_2 \cos{\theta})
\end{array}
\]

All non-trivial integrals over $\theta$ have been performed
numerically.

\newpage
{\footnotesize
$^*$ Present address: Universidade Federal de  S\~ao Carlos,
Dept. F\'{\i}sica, 13565-905 S\~ao Carlos, SP, Brazil. Email:
gregorio@df.ufscar.br }

\newpage


\end{document}